\documentclass[aps,pra,reprint,twocolumn,superscriptaddress,floatfix,nofootinbib,longbibliography]{revtex4-1}
\usepackage{graphicx,amsmath,amsfonts,amssymb,amsthm,xr}
\usepackage{epsfig,amsmath,amssymb,color,dsfont,upgreek,physics}
\usepackage{mathrsfs}
\usepackage{mathtools}
\usepackage{bbold}
\usepackage{comment}
\usepackage{float}

\usepackage[bookmarks=true,colorlinks,linkcolor=OrangeRed,urlcolor=NavyBlue,citecolor=RoyalBlue]{hyperref}
\usepackage[dvipsnames]{xcolor}

\definecolor{mygold}{rgb}{0.93,0.69,0.13}

\definecolor{mypurple}{rgb}{0.49,0.18,0.56}

\definecolor{mygreen}{rgb}{0,0.5,0}

\definecolor{mygreen}{rgb}{0,0.5,0}
\definecolor{myred}{rgb}{0.7,0,0}

\newcommand{\bigzero}{\mbox{\normalfont\Large\bfseries 0}}
\newcommand{\rvline}{\hspace*{-\arraycolsep}\vline\hspace*{-\arraycolsep}}

\usepackage{soul}

\graphicspath{{./Figures/}}

\begin{document}
\title{Prominent quantum many-body scars in a truncated Schwinger model}
\author{Jean-Yves Desaules}
\affiliation{School of Physics and Astronomy, University of Leeds, Leeds LS2 9JT, UK}
\author{Ana Hudomal}
\affiliation{School of Physics and Astronomy, University of Leeds, Leeds LS2 9JT, UK}
\affiliation{Institute of Physics Belgrade, University of Belgrade, 11080 Belgrade, Serbia}
\author{Debasish Banerjee}
\affiliation{Theory Division, Saha Institute of Nuclear Physics, 1/AF Bidhan Nagar, Kolkata 700064, India}
\affiliation{Homi Bhabha National Institute, Training School Complex, Anushaktinagar, Mumbai 400094,India}
\author{Arnab Sen}
\affiliation{School of Physical Sciences, Indian Association for the Cultivation of Science, Kolkata 700032,
India}
\author{Zlatko Papi\'c}
\affiliation{School of Physics and Astronomy, University of Leeds, Leeds LS2 9JT, UK}
\author{Jad C.~Halimeh}
\email{jad.halimeh@physik.lmu.de}
\affiliation{Department of Physics and Arnold Sommerfeld Center for Theoretical Physics (ASC), Ludwig-Maximilians-Universit\"at M\"unchen, Theresienstra\ss e 37, D-80333 M\"unchen, Germany}
\affiliation{Munich Center for Quantum Science and Technology (MCQST), Schellingstra\ss e 4, D-80799 M\"unchen, Germany}

\begin{abstract} 
The high level of control and precision achievable in current synthetic quantum matter setups has enabled first attempts at quantum-simulating various intriguing phenomena in condensed matter physics, including those probing thermalization or its absence in closed quantum systems. In a recent work [Desaules \textit{et al.},~\href{https://arxiv.org/abs/2203.08830}{arXiv:2203.08830}], we have shown that quantum many-body scars---special low-entropy eigenstates that weakly break ergodicity in nonintegrable systems---arise in spin-$S$ quantum link models that converge to $(1+1)-$D lattice quantum electrodynamics (Schwinger model) in the Kogut--Susskind limit $S\to\infty$. In this work, we further demonstrate that quantum many-body scars exist in a truncated version of the Schwinger model, and are qualitatively more prominent than their counterparts in spin-$S$ quantum link models. We illustrate this by, among other things, performing a finite-$S$ scaling analysis that strongly suggests that scarring persists in the truncated Schwinger model in the limit $S\to\infty$. Although it does not asymptotically converge to the Schwinger model, the truncated formulation is relevant to synthetic quantum matter experiments, and also provides fundamental insight into the nature of quantum many-body scars, their connection to lattice gauge theories, and the thermalization dynamics of the latter. Our conclusions can be readily tested in current cold-atom setups.
\end{abstract}

\date{\today}
\maketitle

\tableofcontents

\section{Introduction}
Recent advances in the development of a suite of synthetic quantum matter platforms that possess high levels of control and precision at the single-atom level \cite{Bakr2009} have revolutionized the study of exotic quantum phenomena~\cite{Greiner2002,Bloch2008}.
There is a major ongoing effort in the scientific community to utilize this technological advancement to address various problems on analog and digital quantum simulators \cite{Pasquans_review,Alexeev_review,klco2021standard,aidelsburger2021cold}. Of particular interest in this endeavor are quantum simulations probing the Eigenstate Thermalization Hypothesis (ETH) that shed light on the nature of thermalization or lack thereof in closed quantum systems \cite{Deutsch1991,Srednicki1994,Rigol_2008,Eisert2015,Rigol_review,Deutsch_review}. 
This has led to experimental observations of (pre)thermalization and out-of-equilibrium phase transitions in nonintegrable quantum many-body models \cite{Kaufman2016,Jurcevic2017,Neyenhuis2017,Zhang2017dpt,Kaplan2020,Zhou2021}.

Paradigms of strong ergodicity breaking such as many-body localization (MBL) \cite{Basko2006,Nandkishore_review,Abanin_review} have also received major experimental attention in recent years \cite{Schreiber2015,Smith2016,Kondov2015,Choi2016,Roushan2017,chiaro2020direct,Rispoli2019,Lukin2019}. In MBL systems, sufficiently strong quenched disorder leads to localization in the dynamics of observables over all practically relevant evolution times. This disorder-induced MBL violates ETH in a way not so different from integrable models~\cite{sutherland2004beautiful}. The strong disorder in the system gives rise to an extensive number of local integrals of motion~\cite{Serbyn2013,Huse2014}, which can
indefinitely delay thermalization. However, strong ergodicity breaking can also occur without any disorder, for example when a constant electric field is introduced in a chain of interacting spinless fermions \cite{Schulz2019}. This Stark-MBL has recently been demonstrated experimentally \cite{Morong2021}. Moreover, it has recently been shown that in Stark-MBL systems there exist quasi-localized dynamical l-bits, which are exponentially stable in time and prohibit thermalization \cite{gunawardana2022dynamical}. Yet another example of strong ergodicity breaking has appeared in the context of gauge theories \cite{Brenes2018,Smith2017,Metavitsiadis2017,smith2017absence,Russomanno2020,Papaefstathiou2020,karpov2021disorder,hart2021logarithmic,Zhu2021,Chakraborty2022}, where quenches from an initial state forming a superposition over an extensive number of gauge superselection sectors lead to localized dynamics. This disorder-free localization is caused by an effective disorder emerging over the different background charges of the underlying superselection sectors.

More recently, a new concept of \textit{weak} ergodicity breaking, dubbed quantum many-body scars, has received much attention in the literature~\cite{Serbyn2020, MoudgalyaReview}.
Scarring involves the presence of special nonthermal eigenstates existing at equal energy intervals over the entire spectrum of a nonintegrable, usually disorder-free, interacting model \cite{Moudgalya2018,BernevigEnt, Turner2018, Iadecola2019_2, MotrunichTowers,Zhao2020,Zhao2021}. A characteristic signature of scarred eigenstates is their anomalously low bipartite entanglement entropy, even when they exist in the middle of the spectrum \cite{ShiraishiMori,lin2018exact}. As such, these eigenstates span a ``cold'' subspace, weakly connected to the rest of the Hilbert space \cite{Turner2018}. When a system is prepared in this cold subspace and subsequently quenched, the dynamics will take significantly longer to ``leak out'' into the rest of the Hilbert space, thus delaying the thermalization of the system. This manifests as persistent oscillations in certain local observables, reminiscent of single-particle chaotic systems~\cite{wenwei18TDVPscar}. Scarred dynamics was first observed in a Rydberg-atom system \cite{Bernien2017, Bluvstein2021}. More recently, various signatures of weak ergodicity breaking have also been observed in several other ultracold atom platforms~\cite{Kao2021,Scherg2020,Jepsen2021, Su2022}.

Interestingly, the Ising-type quantum spin model realized in the experiment of Ref.~\cite{Bernien2017} was later shown to map onto the spin-$1/2$ $\mathrm{U}(1)$ quantum link model (QLM), which is a quantum link formulation \cite{Chandrasekharan1997,Wiese_review} of $(1{+}1)-$D quantum electrodynamics on a lattice, known as the lattice Schwinger model. This, along with other works proving the existence of quantum many-body scars in various discrete lattice gauge theories \cite{Banerjee2021,aramthottil2022scar,biswas2022scars} and the necessity of gauge-symmetry stability for their robustness \cite{Halimeh2022robust}, has led to the natural question of whether scars are an inherent feature of ``standard'' lattice gauge theories with a continuous configuration space. In a recent work by us \cite{Desaules2022weak}, we have shown that quantum many-body scars persist at larger link spin lengths $S>1/2$ in the spin-$S$ $\mathrm{U}(1)$ QLM in the form of \textit{resonant scarring} when the initial state is an extreme vacuum, defined as the most highly excited vacuum state. We have additionally presented evidence of \textit{detuned scarring}, recently demonstrated experimentally for $S=1/2$ \cite{Su2022}, also for $S>1/2$ when starting in the physical vacuum or the charge-proliferated state. This all indicates that scarring behavior in the $\mathrm{U}(1)$ QLM is quite rich also beyond $S=1/2$. But is this richness directly related to the specific quantum link formulation that begets the $\mathrm{U}(1)$ QLM from the lattice Schwinger model, or can other, perhaps cruder, representations of the latter also yield equally rich scarring behavior? 

In this work, we address the previous question by comparing and contrasting the spin-$S$ $\mathrm{U}(1)$ QLM with the spin-$S$ \textit{truncated Schwinger model} (TSM), in which the gauge field is a crude truncation of its counterpart in the lattice Schwinger model. We show that the TSM exhibits qualitatively more pronounced scarring behavior than
the spin-$S$ $\mathrm{U}(1)$ QLM. Furthermore, through a finite-$S$ scaling analysis, we conclude that scarring is significantly more likely to persist in the TSM than the QLM in the Kogut--Susskind limit $S\to\infty$.

The rest of the paper is organized as follows: In Sec.~\ref{sec:map}, we first present the mappings used to obtain the two constrained spin models---the spin-$S$ $\mathrm{U}(1)$ QLM and TSM---that we investigate in this work from the lattice Schwinger model, and we showcase their scarring behavior. In Sec.~\ref{sec:comp_PXP}, we show how these models differ from the generalized PXP model studied in Ref.~\cite{wenwei18TDVPscar}, and that they admit a different semiclassical limit.
In Sec.~\ref{sec:S_lim}, we investigate the fate of scarring in the limit of $S\to\infty$, and contrast the TSM and QLM results. 
Finally, in Sec.~\ref{sec:conclusion}, we summarize our findings and discuss their implications. Appendices contain further information and (numerical and analytical) details supporting the main results of the paper.

\section{Spin-$S$ $\mathrm{U}(1)$ quantum link model and truncated Schwinger model}\label{sec:map}
We now consider the spin-$S$ $\mathrm{U}(1)$ QLM and the TSM, both of which are directly relevant to modern experiments in synthetic quantum matter probing quantum electrodynamics. The lattice Schwinger model, which is quantum electrodynamics in $(1+1)$ dimensions, includes gauge fields with an infinite-dimensional local Hilbert space. Whereas the spin-$S$ $\mathrm{U}(1)$ QLM substitutes these operators with spin-$S$ matrices, retrieving the lattice Schwinger model asymptotically in the limit $S\to\infty$, the TSM involves explicitly truncating these operators. In contrast to the case of the QLM, this forbids an asymptotic approach to the lattice Schwinger model as $S\to\infty$ in the case of the TSM. Nevertheless, both the QLM and TSM are lattice gauge theories, and can therefore provide deep insights into the connection between the underlying gauge symmetry and the corresponding scarring behavior.

\subsection{Mapping to a constrained spin model}
Let us now take advantage of the gauge symmetry of the spin-$S$ $\mathrm{U}(1)$ QLM and the TSM in order to map them into spin models that are invariant under translation. To achieve this, we follow a procedure similar to the one used in Ref.~\cite{Surace2020} to map to the PXP model. We start in the lattice Schwinger model, described by the Hamiltonian
\begin{align}\nonumber
    \hat{H}&=\frac{J}{2a}\sum_{j=1}^L\left(\hat{\phi}_j^\dagger\hat{U}_{j,j+1}\hat{\phi}_{j+1}+\mathrm{H.c.}\right)\\\label{eq:H_KS}
    &+\mu\sum_{j=1}^L (-1)^j\hat{\phi}_j^\dagger\hat{\phi}_j+\frac{a}{2}\sum_{j=1}^L \left(\hat{E}_{j,j+1}\right)^2,
\end{align}
where $a$ is the lattice spacing and $\mu$ is the fermionic mass. We henceforth set $a=1$ throughout this work. Matter fields are represented by the staggered fermionic operators $\hat{\phi}_j^{(\dagger)}$ at lattice site $j$, while the gauge and electric fields $\hat{U}_{j,j+1}$ and $\hat{E}_{j,j+1}$ reside on the link between lattice sites $j$ and $j+1$, and satisfy the commutation relations
\begin{subequations}\label{eq:comm}
\begin{align}\label{eq:comm1}
    \big[\hat{U}_{j,j+1},\hat{U}_{l,l+1}^\dagger\big]&=0,\\\label{eq:comm2}
    \big[\hat{E}_{j,j+1},\hat{U}_{l,l+1}\big]&=g\delta_{j,l}\hat{U}_{j,j+1}.
\end{align}
\end{subequations}
The lattice Schwinger model is a $\mathrm{U}(1)$ gauge theory, and Gauss's law is given by $\hat{G}_j\ket{\text{phys}}=0,\,\forall j$, where $\ket{\text{phys}}$ is a physical state and the local gauge-symmetry generator is
\begin{equation}
    \hat{G}_j=\hat{E}_{j,j+1}-\hat{E}_{j-1,j}-g\left[\hat{\phi}_j^\dagger\hat{\phi}_j-\frac{1-(-1)^j}{2}\right],
\end{equation}
and $g$ is the gauge coupling. 

In the quantum link formulation, the gauge and electric fields are represented by finite-dimensional spin-$S$ operators:
\begin{subequations}\label{eq:trafo}
\begin{align}
    \hat{U}_{j,j+1}&\to\frac{\hat{s}^+_{j,j+1}}{\sqrt{S(S+1)}},\\
    \hat{E}_{j,j+1}&\to g\hat{s}_{j,j+1}^z,
\end{align}
\end{subequations}
where the factor $1/\sqrt{S(S+1)}$ ensures the correct scaling with $S$ \cite{kasper2017}. This can be checked by substituting~\eqref{eq:trafo} into the commutation relations~\eqref{eq:comm} to get:
\begin{subequations}
\begin{align}\nonumber
    \big[\hat{U}_{j,j+1},\hat{U}_{l,l+1}^\dagger\big]&\to\frac{1}{S(S+1)}\big[\hat{s}^+_{j,j+1},\hat{s}^-_{l,l+1}\big]\\\label{eq:comm1p}
    &=\frac{2\delta_{j,l}}{S(S+1)}\hat{s}^z_{j,j+1},\\\nonumber
    \big[\hat{E}_{j,j+1},\hat{U}_{l,l+1}\big]&\to \frac{g}{\sqrt{S(S+1)}}\big[\hat{s}^z_{j,j+1},\hat{s}^+_{j,j+1}\big]\\\label{eq:comm2p}
    &=g\delta_{j,l}\frac{\hat{s}^+_{j,j+1}}{\sqrt{S(S+1)}}.
\end{align}
\end{subequations}
Whereas Eq.~\eqref{eq:comm2} is automatically satisfied at any $S$ through Eq.~\eqref{eq:comm2p}, we find that Eq.~\eqref{eq:comm1} is achieved through Eq.~\eqref{eq:comm1p} asymptotically in the limit $S\to\infty$.

In order to obtain a Hamiltonian that is invariant under translation, we directly incorporate the particle-hole transformation 
\begin{align}
\hat{\phi}_j\to \frac{1+(-1)^j}{2}\hat{\phi}_j+\frac{1-(-1)^j}{2}\hat{\phi}_j^\dagger.    
\end{align}
This has the consequence of taking
\begin{align}\label{eq:fermionPH}
\hat{\phi}_j^\dagger\hat{\phi}_j\to \frac{1-(-1)^j}{2}+(-1)^j\hat{\phi}_j^\dagger\hat{\phi}_j,  
\end{align}
due to the fermionic anticommutation relations. For our field operators, this particle-hole transformation takes the form
\begin{subequations}\label{eq:phq}
\begin{align} 
    \hat{U}_{j,j+1}&\to\frac{1-(-1)^j}{2\sqrt{S(S+1)}}\hat{s}^+_{j,j+1}{+}\frac{1+(-1)^j}{2\sqrt{S(S+1)}}\hat{s}^-_{j,j+1},\\
    \hat{E}_{j,j+1}&\to g(-1)^{j+1}\hat{s}_{j,j+1}^z.
\end{align}
\end{subequations}
rendering Hamiltonian~\eqref{eq:H_KS} in the form
\begin{align}\nonumber
    \hat{H}&=\frac{J}{2\sqrt{S(S+1)}}\sum_{j=1}^L\left(\hat{\phi}_j\hat{s}^+_{j,j+1}\hat{\phi}_{j+1}+\mathrm{H.c.}\right)\\
    &+\mu\sum_{j=1}^L \hat{\phi}_j^\dagger\hat{\phi}_j+\frac{g^2}{2}\sum_{j=1}^L \left(\hat{s}^z_{j,j+1}\right)^2.
\end{align}
The generator of Gauss's law can now be rewritten in the simple form
\begin{align}
    \hat{G}_j=(-1)^{j+1} g \left(\hat{s}^z_{j,j+1}+\hat{s}^z_{j-1,j}+\hat{\phi}_j^\dagger\hat{\phi}_j\right).
\end{align}
As $\hat{\phi}_j^\dagger\hat{\phi}_j$ can only be equal to zero or $\mathds{1}$, this means that $\hat{s}^z_{j,j+1}+\hat{s}^z_{j-1,j}$ must always be equal to zero or $-\mathds{1}$, respectively, in the physical sector $\hat{G}_j\ket{\text{phys}}=0,\,\forall j$.
Restricting to this sector and employing periodic boundary conditions (PBC), we can therefore replace the mass term by $-\mu\sum_j\big( \hat{s}^z_{j,j+1}+\hat{s}^z_{j-1,j}\big)=-2\mu\sum_j \hat{s}^z_{j,j+1}$.
By then integrating out the fermionic degrees of freedom, we end up with the translation-invariant Hamiltonian 
\begin{align}\label{eq:H_spin}
    \hat{H}_\mathrm{QLM}=J_S\,\hat{\mathcal{P}}\bigg(\sum_j \hat{s}^x_j\bigg)\hat{\mathcal{P}}
    -2\mu\sum_j \hat{s}^z_j+\frac{g^2}{2}\sum_j\left(\hat{s}_j^z\right)^2,
\end{align}
where $\hat{\mathcal{P}}$ is a projector that annihilates all states outside of the physical sector, and $J_S=J/\sqrt{S(S+1)}$.

However, this is not the only way to obtain a finite-dimensional model for which we recover the lattice Schwinger model at infinite $S$. One can also replace $\hat{U}_{j,j+1}$ in Eq.~\eqref{eq:H_KS} not by a spin-$S$ operator but by the $(2S+1)\times(2S+1)-$dimensional operator $\hat{\tau}^+_{j,j+1}$. This operator has the same structure as the $\hat{s}^+_{j,j+1}$, with the exception that each of the latter's nonzero matrix elements is replaced by $1$. In other words, $\hat{\tau}^+_{j,j+1}$ is merely the $(2S+1)\times(2S+1)-$dimensional truncated version of $\hat{U}_{j,j+1}$ at its center. The field operator can then be represented by the matrix $\hat{\tau}^z_{j,j+1}=\hat{s}^z_{j,j+1}$, and we thus identify our fields as
\begin{subequations}\label{eq:trunc}
\begin{align}
    \hat{U}_{j,j+1}&\to\hat{\tau}^+_{j,j+1},\\
    \hat{E}_{j,j+1}&\to g\hat{\tau}_{j,j+1}^z.
\end{align}
\end{subequations}
The commutation relations~\eqref{eq:comm} then become
\begin{subequations}
\begin{align}\nonumber
    \big[\hat{U}_{j,j+1},\hat{U}_{l,l+1}^\dagger\big]&\to\big[\hat{\tau}^+_{j,j+1},\hat{\tau}^-_{l,l+1}\big]=\delta_{j,l}\hat{\Sigma}_{j,j+1}\\\label{eq:comm1t}
    &=\delta_{j,l}\begin{pmatrix} 1 & \rvline &\begin{matrix} 0 & \cdots &  0\end{matrix} & \rvline & 0\\
\hline
  \begin{matrix} 0\\[-0.4em] \vdots\\[-0.4em]  0\end{matrix}  & \rvline & \bigzero & \rvline & \begin{matrix}0\\[-0.4em] \vdots\\[-0.4em]  0\end{matrix}   \\ 
  \hline
  0 & \rvline & \begin{matrix} 0 & \cdots &  0\end{matrix}  & \rvline & -1
\end{pmatrix}_{j,j+1},\\
    \big[\hat{E}_{j,j+1},\hat{U}_{l,l+1}\big]&\to g\big[\hat{\tau}^z_{j,j+1},\hat{\tau}^+_{j,j+1}\big]=g\delta_{j,l}\hat{\tau}^+_{j,j+1}. \label{eq:comm2t}
\end{align}
\end{subequations}
Similarly to the case of the QLM, the commutation relation~\eqref{eq:comm2} is satisfied through Eq.~\eqref{eq:comm2t} for any $S$. However, Eq.~\eqref{eq:comm1} is satisfied through Eq.~\eqref{eq:comm1t} only strictly at infinite $S$, rather than approach it asymptotically as in the case of the QLM. Indeed, the commutation relation~\eqref{eq:comm1t} is always equal to $\hat{\Sigma}_{j,j+1}$, which is a $(2S+1)\times(2S+1)$ matrix with zeros everywhere except in entries $(1,1)$ and $(2S+1,2S+1)$, which are respectively plus and minus unity.

By employing the particle-hole transformations~\eqref{eq:fermionPH} to set
\begin{subequations}\label{eq:phq2}
\begin{align} 
    \hat{U}_{j,j+1}&\to\frac{1-(-1)^j}{2}\hat{\tau}^+_{j,j+1}{+}\frac{1+(-1)^j}{2}\hat{\tau}^-_{j,j+1},\\
    \hat{E}_{j,j+1}&\to g(-1)^{j+1}\hat{\tau}_{j,j+1}^z,
\end{align}
\end{subequations}
and utilizing Gauss's law to integrate out the fermionic degrees of freedom as done before in the case of the QLM, we obtain
\begin{align}\label{eq:H_TSM}
    \hat{H}_\mathrm{TSM}=\hat{\mathcal{P}}\bigg(\sum_j \hat{\tau}^x_j\bigg)\hat{\mathcal{P}}
    -2\mu\sum_j \hat{\tau}^z_j+\frac{g^2}{2}\sum_j\left(\hat{\tau}_j^z\right)^2,
\end{align}
where $\hat{\tau}^x_j=(\hat{\tau}^+_j+\hat{\tau}^-_j)/2$ is the tridiagonal matrix with entries of $0$ along the principal diagonal and entries of $1/2$ along both the sub- and superdiagonal. Note that as $\hat{\tau}^z_{j,j+1}=\hat{s}^z_{j,j+1}$, Gauss's law goes through the same transformation as for the QLM, and the global projector $\hat{\mathcal{P}}$ is the same in the two models. As a consequence, the Hilbert space is also the same. However due to the difference in the matrix elements of the Hamiltonian, the two models are only equivalent (up to an overall multiplicative factor) for $S=1/2$ and $S=1$, which are the only cases where the spin operator $\hat{s}^x_j$ has equal matrix elements.

The difference on how the TSM and QLM approach the Kogut--Susskind limit ($S\to\infty$) can be highlighted by looking at the 2-norms of the commutators~\eqref{eq:comm1p} and~\eqref{eq:comm1t}. As both operators are diagonal, the 2-norm is simply the largest absolute value of any diagonal entry. In the case of the TSM, the commutator in Eq.~\eqref{eq:comm1t} will always have a 2-norm of unity for any finite $S$ no matter how large $S$ is. In the case of the QLM, the commutator of Eq.~\eqref{eq:comm1p} has a 2-norm of $2/(S+1)$. Consequently, the QLM asymptotically approaches the lattice Schwinger model as $S\to\infty$, while the TSM does not. We emphasize that this result does not depend on the choice of the norm. As an example, even if we choose the Frobenius norm we get in the case of the TSM a norm of $\sqrt{2}$ at any $S$, while in the case of the QLM it is equal to $2\sqrt{(2S+1)/[3S(S+1)]}$. This is the main reason why from a gauge-theory perspective the QLM is favored over the TSM, as one can employ the QLM and controllably study the approach to the Kogut--Susskind limit of quantum electrodynamics. Nevertheless, it is important to emphasize that here we are interested in scarring behavior, and as we will show in this work, the TSM has more prominent scars than the QLM. Therefore, from a quantum many-body scars perspective and given its experimental feasibility, the TSM is fundamentally as relevant for our purposes as the QLM is.

For $S=1/2$, both Eqs.~\eqref{eq:H_spin} and \eqref{eq:H_TSM} reduce to the well-known PXP model~\cite{FendleySachdev,Lesanovsky2012} (although with a slightly different prefactor in front). 
However, we stress that for $S>1/2$ these models are different from the recently developed generalized PXP model \cite{wenwei18TDVPscar}, and we address these differences in Sec.~\ref{sec:comp_PXP}.

Before delving into the properties of the QLM and TSM, we mention a few states that have physical significance for any value of $S$.  The first one is the vacuum state (i.e., with no matter present in the Schwinger model), with the highest possible value of the electric field, which we shall henceforth refer to as the \emph{extreme vacuum}. This state corresponds to the ground state of the TSM or QLM with $\mu\to\infty$ and $g^2<0$. Correspondingly, we will denote it by $\ket{0_-}$, with the subscript denoting the fact that $g^2$ is negative. This state is doubly degenerate for any value of $S$, and, working in the basis of the $\hat{s}^z$, we set $\ket{0_-}=\ket{S,-S,\ldots,S,-S}$ and $\ket{0_-^{\prime}}=\ket{-S,S,\ldots,-S,S}$. 
For a more physical set of parameters, we can have the ground state at $\mu\to\infty$ but with $g^2$ positive. We call this state the \emph{physical vacuum} and denote it by $\ket{0_+}$. The physical vacuum state is $\ket{0_+}=\ket{0,0\ldots 0,0,0}$ for integer $S$. For half-integer $S$, it is doubly degenerate and given by $\ket{0_+}=\ket{1/2, -1/2,\ldots,1/2,-1/2}$ and $\ket{0_+^{\prime}}=\ket{-1/2, 1/2,\ldots,-1/2,1/2}$. In this formulation, it is immediately clear that for the PXP case (i.e., $S=1/2$) the extreme and physical vacua are identical, which is expected as the electric field coupling term is just an inconsequential energetic constant. We also are interested in the ground state of the TSM and QLM for $\mu\to-\infty$ and $g^2$ positive. In that case the presence of fermions is energetically favorable, and so the ground state is the \emph{charge-proliferated state} with the maximal matter occupation, which we denote by $\ket{{\rm CP}}$. For integer $S$, this state is doubly degenerate with the two states being $\ket{{\rm CP}}=\ket{0,-1,\ldots,0,-1}$ and $\ket{{\rm CP}^{\prime}}=\ket{-1,0,\ldots -1,0}$. For half-integer $S$, it is nondegenerate and given by $\ket{{\rm CP}}=\ket{-1/2, -1/2,\ldots,-1/2,-1/2}$. This state corresponds to the polarized state in the case of the PXP model.
\begin{figure}[t!]
\centering
\includegraphics[width=\linewidth]{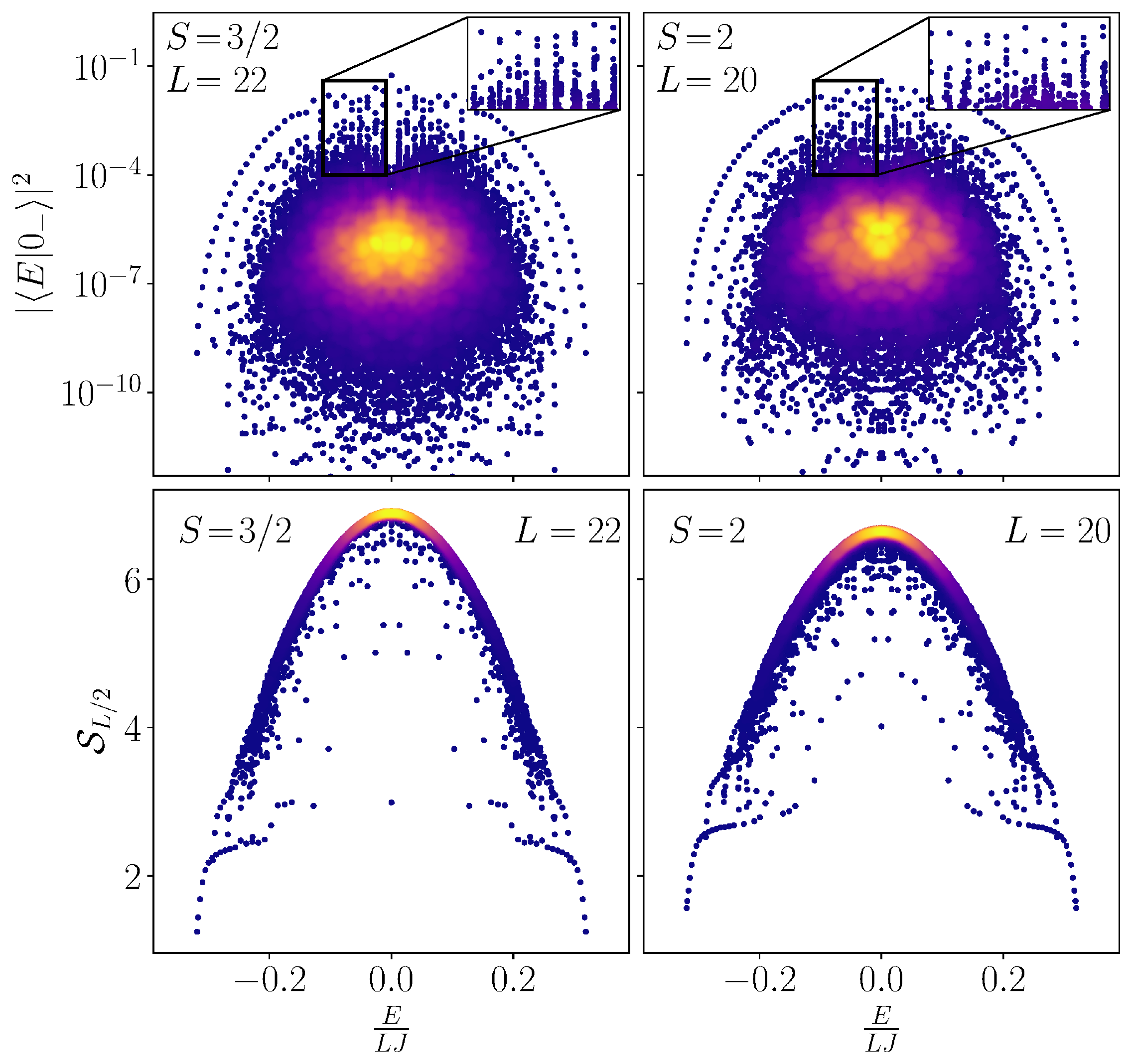}
\caption{Overlap between the extreme vacuum $\ket{0_-}$ and the eigenstates (top) and entanglement entropy of the eigenstates (bottom) in the spin-$S$ TSM at $\mu=g=0$. The color indicates the density of data points (the brighter the denser). The top band of scarred states is clearly visible for all $S$, and the states in it have anomalously low entanglement entropy.}
\label{fig:LGT_eq_olap}
\end{figure}

\begin{figure}[t!]
\centering
\includegraphics[width=\linewidth]{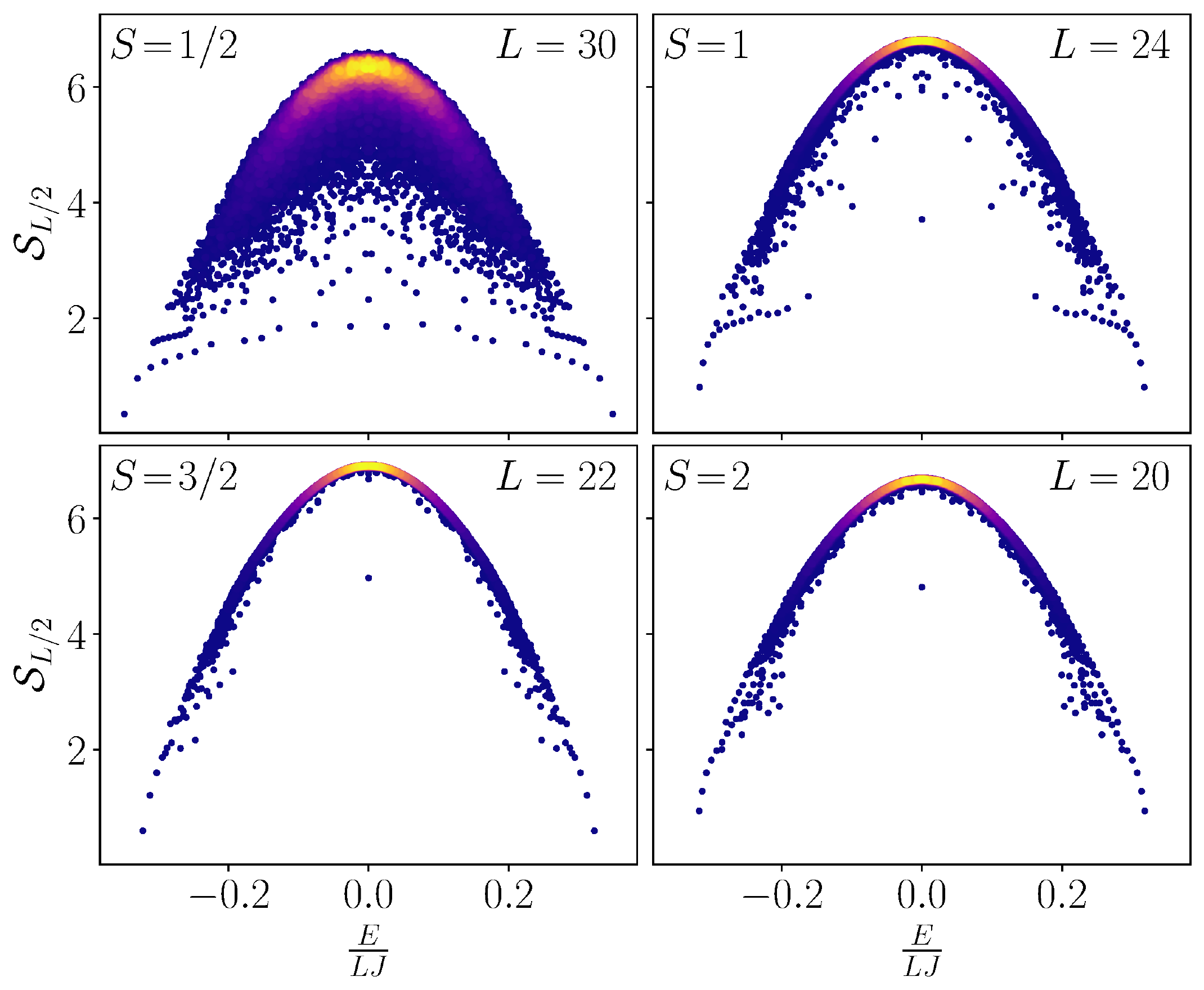}
\caption{Entanglement entropy of the eigenstates of the QLM for $S=1/2$ to $S=2$. The color indicates the density of data points. The scarred eigenstates away from the edges of the spectrum have an entanglement entropy close to that of the bulk, except for a single zero-mode.}
\label{fig:LGT_S_eig}
\end{figure}

\subsection{Resonant scarring: quenches to $\mu=g=0$}\label{sec:resonant}
In the spin-$1/2$ QLM, the scarred states are the two vacua $\ket{0_-}=\ket{0_+}$ and $\ket{0_-^{\prime}}=\ket{0_+^{\prime}}$ for the quench Hamiltonian with $\mu=g=0$ (resonant scarring). Numerical simulations show that for any $S>1/2$ only the extreme vacua $\ket{0_-}$ and $\ket{0_-^{\prime}}$ exhibit scarring \cite{Desaules2022weak} (see Appendix~\ref{app:other_states} for results on the physical vacuum $\ket{0_+}$). A typical signature of scarring is a strong overlap of the scarred state with a small set of eigenstates that are approximately equally spaced in energy. We have demonstrated this for the spin-$S$ $\mathrm{U}(1)$ QLM in Ref.~\cite{Desaules2022weak} for the extreme vacua. However, this can be seen even more clearly for the TSM, as shown in Fig.~\ref{fig:LGT_eq_olap} for the overlap of the extreme vacuum with the eigenstates of the TSM at $\mu=g=0$ for $S=3/2$ and $2$. A top band with $2SL+1$ states is well-separated from the bulk of states. Note that for $S=1/2$ and $1$, the TSM and QLM are identical, which is why we do not show results for these values of $S$ as they can already be found in Ref.~\cite{Desaules2022weak}.

The band of high-overlap eigenstates is then the set of nonthermal eigenstates that lead to scarring when an initial state is prepared in their subspace. These eigenstates are expected to exhibit nonthermal properties such as anomalously low bipartite entanglement entropy. Indeed, this is also observed for the TSM especially near the edges of the spectrum, as shown in Fig.~\ref{fig:LGT_eq_olap}.
It is worth noting that the picture is much less clear in the case of the QLM at $S>1$, where only a single zero-mode exhibits anomalously low entanglement entropy, as shown in Fig.~\ref{fig:LGT_S_eig}. This is due to the ``dilution'' of the scarring among many eigenstates. Indeed, the overlap between $\ket{0_-}$ and the eigenstates does not really show a single well-separated band of states with high overlap in the case of the QLM, but rather a set of $2SL+1$ towers rising above the bulk of states. So instead of having one clear scarred eigenstate per tower with highly atypical behavior, we have a larger number of eigenstates sharing this atypicality. This is similar to what is observed in the PXP model when scarred eigenstates hybridize with thermal states. As a consequence, the former becomes more thermal while the latter display atypical properties. As a result, near the middle of the spectrum in the QLM there is no single eigenstate showing very low entanglement entropy. Instead, there are several outliers just below the band of thermal states. This kind of phenomenology has already been seen in other models with inexact scars \cite{DesaulesTFH,zhang2022}.
As $S$ is increased, the towers seem to get denser, without any clear state at the top even relatively close to the edges of the spectrum. As a consequence the outliers in entanglement entropy get increasingly closer to the value of thermal states in the case of the QLM, in contrast to the case of the TSM, where both in the overlap and entanglement entropy, the scarred eigenstates are clearly separated from their thermal counterparts.
The only exception for the QLM is the presence of a single low-entanglement zero mode. Due to the large number of exactly degenerate zero modes, it may be expected that low-entropy states can be obtained from appropriate linear combinations as demonstrated in  Refs.~\cite{Banerjee2021,Karle2021}.

\begin{figure}[t!]
\centering
\includegraphics[width=\linewidth]{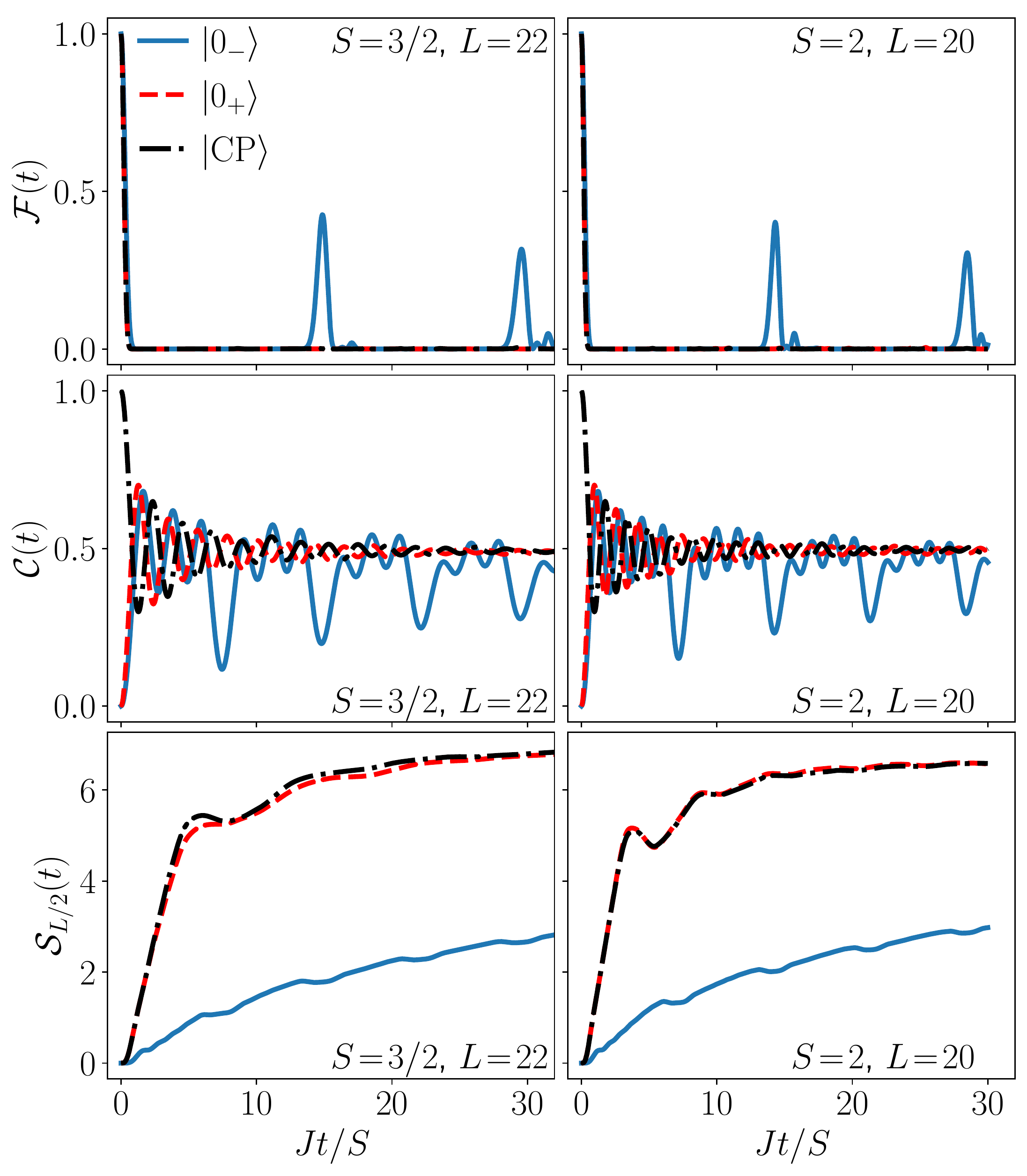}
\caption{Dynamics of the fidelity, chiral condensate, and bipartite entanglement entropy for the TSM with $S=3/2$  and $S=2$ after a zero-mass quench. For all of these quantities the extreme vacuum $\ket{0_-}$ shows anomalous dynamics whereas $\ket{0_+}$ and $\ket{{\rm CP}}$ are thermalizing as expected.}
\label{fig:LGT_eq_dyn}
\end{figure}

In Ref.~\cite{Desaules2022weak}, we have shown that zero-mass zero-$g$ quenches starting in $\ket{0_-}$ lead to revivals in both the fidelity and chiral condensate, 
\begin{subequations}
\begin{align}
    \mathcal{F}(t)&=\big\lvert\braket{\psi_0}{\psi(t)}\big\rvert^2,\\
    \mathcal{C}(t)&=-\frac{2}{L}\sum_{j=1}^L\bra{\psi(t)}\hat{s}^z_j\ket{\psi(t)},
\end{align}
\end{subequations}
respectively, as well as to an anomalously low and slowly growing entanglement entropy. The chiral condensate is a measure of the spontaneous breaking of the chiral symmetry corresponding to fermions in the model by the ensuing dynamics. Here, $\ket{\psi_0}$ is the initial state ($\ket{0_-}$ in this case) and $\ket{\psi(t)}=e^{-i\hat{H}t}\ket{\psi_0}$ where $\hat{H}$ is the quench Hamiltonian. The same behavior occurs also in the TSM for the same quench, as shown in Fig.~\ref{fig:LGT_eq_dyn}, proving that this model exhibits the same scarring phenomenology in the dynamics, as well. Just as in the case of the QLM, scarring dynamics occurs only when starting in $\ket{0_-}$ when the quench Hamiltonian is that of the TSM~\eqref{eq:H_TSM} at $\mu=g=0$. This is evident in the prominence of revivals in the fidelity when starting in $\ket{0_-}$, while starting in $\ket{0_+}$ or $\ket{\text{CP}}$ leads to a fidelity exhibiting fast decay without revivals, which is typical of a thermalizing system. Quantitatively, we find better revivals in the case of the TSM than the QLM~\cite{Desaules2022weak} when starting in $\ket{0_-}$. This is quite surprising, because unlike a free-spin model $\hat{H}=\hat{s}^x$, the unconstrained version of the TSM $\hat{H}=\hat{\tau}^x$ does not lead to perfect revivals for larger values of $S$. Indeed, the latter can simply be mapped to a single particle with uniform hopping on a one-dimensional chain with $2S+1$ states, for which it has been proven that perfect state transfer or revivals are not possible beyond $S=1$ when starting at one end of the chain \cite{Christandl2004}.

For the TSM, we also find a different scaling of the revival period $T_\text{TSM}$ with $S$ than in the case of the QLM. Indeed, unlike in the QLM where we observed numerically that $T_\text{QLM}\approx5.13\pi S$ scales linearly with $S$ \cite{Desaules2022weak}, in the TSM the scaling is actually closer to $\sqrt{S(S+1)}$. In particular, we find that the simplest approximation of the revival period is $T_\mathrm{TSM}\approx 11.5\sqrt{S(S+1)}/J$, which is relatively accurate for smaller $S$. As in the case of the QLM, this revival frequency is set by the energy spacing of the scarred states near the middle of the spectrum.

Turning to local observables such as the chiral condensate, we find persistent oscillations in the latter over all investigated evolution times when the quench starts in $\ket{0_-}$, but not in $\ket{0_+}$ and $\ket{\text{CP}}$ where the oscillations are quickly damped, indicative of thermalization. Also in the case of $\ket{0_-}$ as initial state, we find quantitatively stronger oscillations in the case of the TSM compared to the QLM~\cite{Desaules2022weak}. The frequency of the main oscillations for $\mathcal{C}(t)$ is twice that of $\mathcal{F}(t)$. The reason behind this is that the dynamics after a quench from $\ket{0_-}$ consists in a series of state transfer to $\ket{0_-^{\prime}}$ and back. So after one full revival period the state comes back close to $\ket{0_-}$, but at half of that period it is close to $\ket{0_-^{\prime}}$. The fidelity revival is only high in the first case, but as both states are vacua, the chiral condensate value is close to zero at both times. Smaller oscillations at a faster frequency can also be seen; see Sec.~\ref{sec:S_lim} for more details.

Further confirming this picture of scarring for zero-mass zero-$g$ quenches in the TSM starting in $\ket{0_-}$, we find an anomalously low and slowly growing entanglement entropy of the time-evolved wave function, in sharp contrast to the same quench starting in $\ket{0_+}$ or $\ket{\text{CP}}$ where $\mathcal{S}_{L/2}(t)$ immediately shows fast growth typical of ergodic dynamics. In agreement with our observations for the fidelity and chiral condensate, when starting in $\ket{0_-}$ and quenching with the TSM Hamiltonian~\eqref{eq:H_TSM} at $\mu=g=0$, we find that the entanglement entropy is even lower than for the corresponding quench in the case of the QLM~\cite{Desaules2022weak}.

Therefore, we have shown that the TSM exhibits scarred dynamics very similar to that of the QLM, but with two major differences: (i) the scar eigenstates of the quench Hamiltonian show qualitatively more nonthermal behavior in the overlap with the extreme vacuum, and in the bipartite entanglement entropy (see Fig.~\ref{fig:LGT_eq_olap}), and (ii) scarring dynamics is quantitatively more prominent in the TSM compared to the QLM as can be seen in stronger fidelity revivals, larger chiral condensate oscillations, and a lower entanglement entropy in the former over all investigated evolution times when starting in the extreme vacuum. We will further compare the prominence of scarring between the QLM and TSM in Secs.~\ref{sec:comp_PXP} and~\ref{sec:S_lim}.

\subsection{Detuned scarring: quenches to finite values of $\mu$ and $g^2$}\label{sec:mu_g_quench}
For $\mu=g=0$, only the extreme vacua are scarred (see Appendix~\ref{app:other_states}). However, as in the case of the spin-$1/2$ PXP model, quenching to some specific values of $\mu$ and $g$ can lead to nontrivial athermal behavior when starting in the physical vacuum or the charge-proliferated state \cite{Su2022}. In this section we characterize this behavior for a wider range of these parameters in the QLM.

\begin{figure}[t!]
\centering
\includegraphics[width=\linewidth]{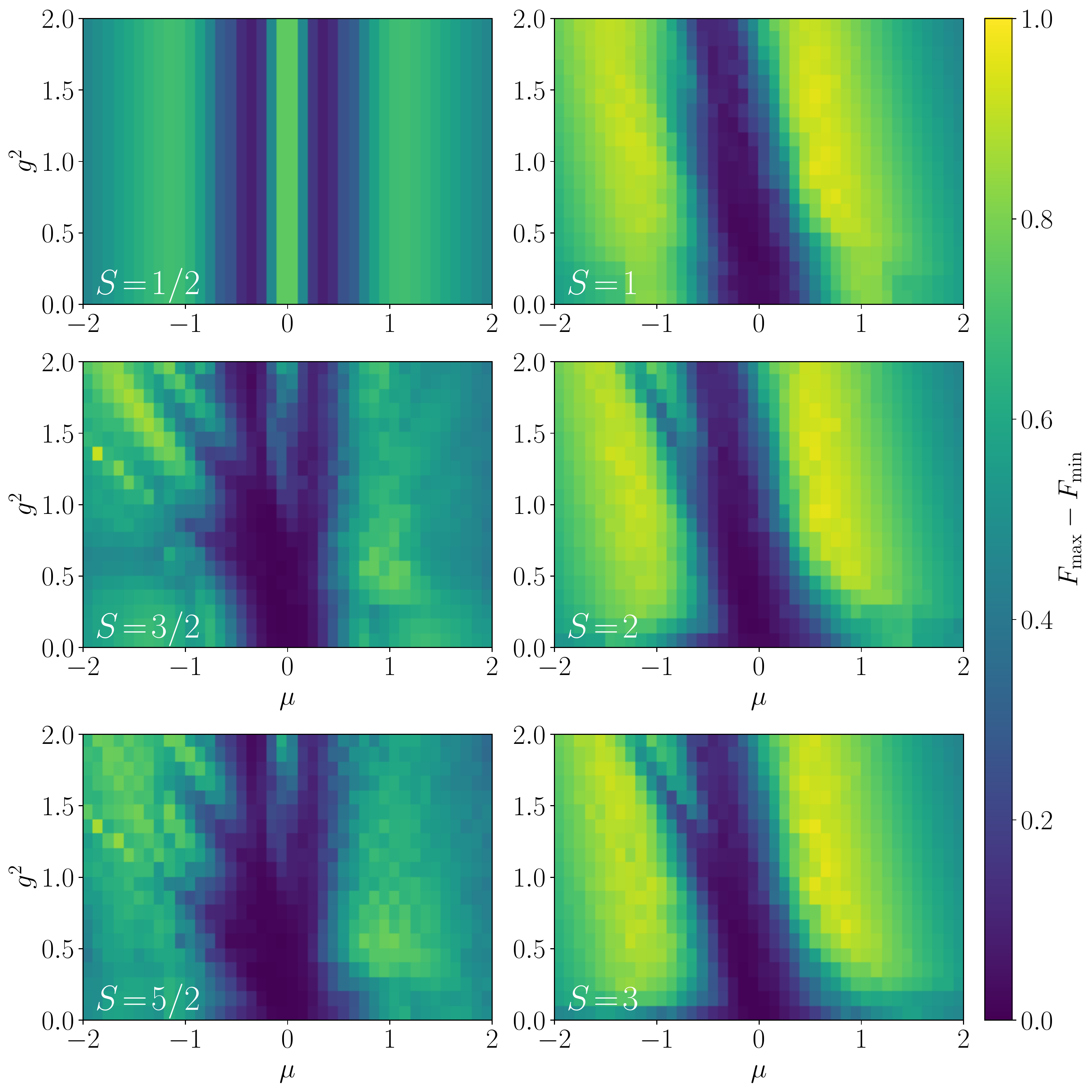}
\caption{Difference between the maximum and minimum amplitudes of the fidelity revival when quenching from the state $\ket{0_+}$ for $L=16$ and various values of $\mu$ and $g^2$ in the spin-$S$ QLM. The revival pattern is different from half-integer and integer $S$, but in both cases we find regions with clear revivals.
}
\label{fig:LGT_quench_0p}
\end{figure}

\begin{figure}[t!]
\centering
\includegraphics[width=\linewidth]{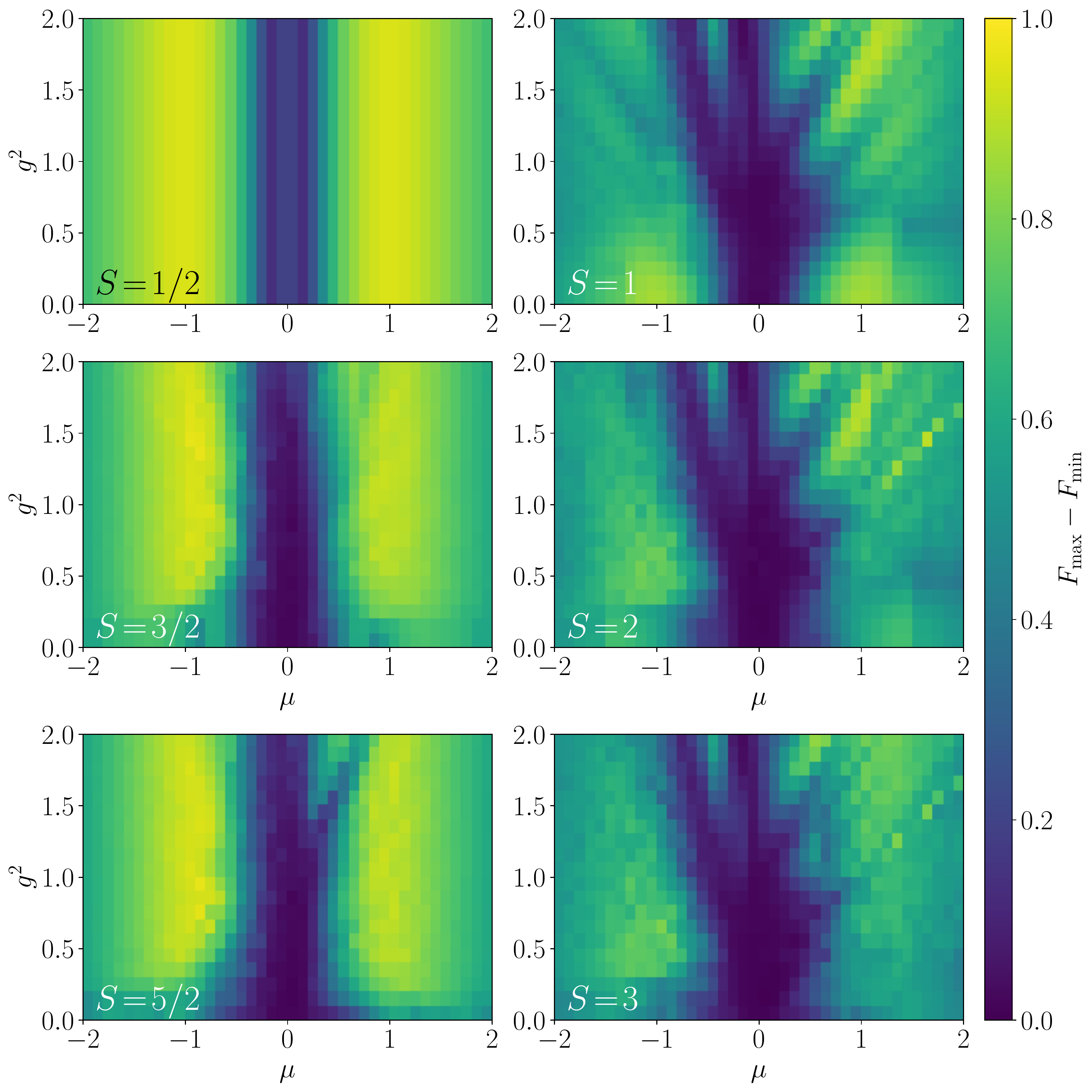}
\caption{Difference between the maximum and minimum amplitudes of the fidelity revival when quenching from the state $\ket{\rm CP}$ for $L=16$ and various values of $\mu$ and $g^2$ in the QLM. For half-integer spin, we find similar detuned scarring for all $S$ as was observed in the spin-$1/2$ PXP model \cite{Su2022}.
}
\label{fig:LGT_quench_CP}
\end{figure}

We study the difference between the maximal and minimal fidelity revival amplitude after a quench for the states $\ket{0_-}$, $\ket{0_+}$, and $\ket{{\rm CP}}$. A large difference indicates good revivals separated by intervals of low fidelity, as is typical when scarring is present. For the $\ket{0_-}$ state, we do see revivals for finite values of $\mu$ and $g^2$, however they are due to Hilbert space fragmentation at the resonance point $\mu=g^2(2S-1)/4$, \ $\mu,g^2\gg J/\sqrt{S}$ (see Appendix \ref{app:det_0m}).
For the $\ket{0_+}$ and $\ket{{\rm CP}}$ states, the pattern is relatively similar but their behavior is switched for integer and half-integer $S$ (Figs.~\ref{fig:LGT_quench_0p} and~\ref{fig:LGT_quench_CP}). The best revivals occur for integer $S$ in the case of $\ket{0_+}$ and for half-integer $S$ in the case of $\ket{{\rm CP}}$. This corresponds to the case where the state is not degenerate.

\begin{figure}[t!]
\centering
\includegraphics[width=\linewidth]{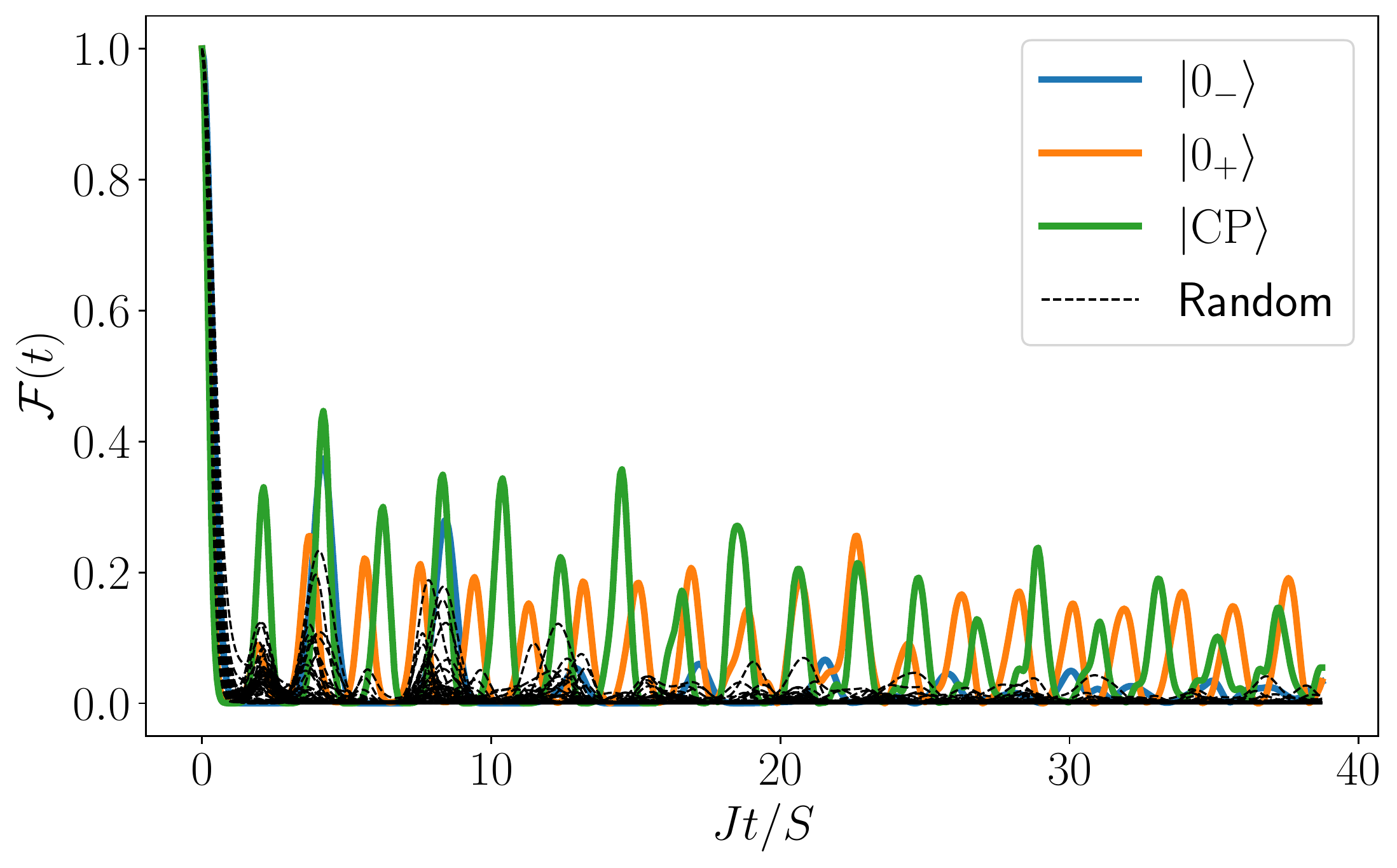}
\caption{Fidelity dynamics for quenches in the QLM starting in the $\ket{0_-}$, $\ket{0_+}$, and $\ket{{\rm CP}}$ states as well as from random basis states for $L=20$, $S=3/2$, $\mu=0.486J$, and $g^2=0.6J$. While several states show some oscillations at short times, only $\ket{0_+}$ and $\ket{{\rm CP}}$ show this behavior persisting at a longer timescale, indicative of detuned scarring.}
\label{fig:LGT_quench_all_D4}
\end{figure}

In Ref.~\cite{Desaules2022weak}, we show an example of a quench leading to detuned scarring in a system with $L=20$  and $S=3/2$. Both states investigated, $\ket{0_+}$ and $\ket{\text{CP}}$ show similar behavior, with the presence of towers of states in the overlap plots and clear revivals in the fidelity with a very slow decay. It is important to note that there are revivals from these two states in a regime where all terms in the Hamiltonian are of equal strength. These oscillations do not have a trivial explanation in terms of conservation of mass or electric energy. The nongenerality of this behavior is highlighted in Fig.~\ref{fig:LGT_quench_all_D4}, which shows quenches from random basis states. For all states shown except $\ket{0_+}$ and $\ket{\text{CP}}$, revivals are either decaying rapidly or not present at all, as expected from a thermalizing system.

\section{Comparison with the spin-$S$ PXP model} \label{sec:comp_PXP}
In Sec.~\ref{sec:map}, we have shown that both the QLM and TSM display scarring for zero-mass zero-$g$ quenches starting in the extreme vacuum $\ket{0_-}$. This is the same state showing scarring in the paradigmatic PXP model at $S=1/2$~\cite{Turner2018b}. For $S>1/2$, the PXP model has been generalized as~\cite{wenwei18TDVPscar}
\begin{align}\label{eq:PXP}
\hat{H}_\mathrm{PXP}=\Omega\sum_j \hat{\mathcal{P}}^{-S}_{j-1}\hat{s}^x_j\hat{\mathcal{P}}^{-S}_{j+1},
\end{align}
with $\hat{\mathcal{P}}^{-S}_j$ the local projector on the lowest possible spin state of eigenvalue $-S$. This model has been studied in Ref.~\cite{wenwei18TDVPscar} for $S>1/2$, and revivals have been found in this case. 

Here, we show that the generalized PXP model~\eqref{eq:PXP} is fundamentally different from the QLM and TSM because of the form of their respective constraints, which results in different Hilbert spaces, and that, as a consequence, their classical limits are also different.

\subsection{Hilbert space structure}

In order to characterize the constraint in the TSM and QLM, it is informative to look at the asymptotic quantum dimension. This quantity tells us how fast the Hilbert space dimension $\mathcal{D}$ grows with the number of sites $L$ as $\mathcal{D}=\alpha^L$, with $\alpha$ the quantum dimension. For an unconstrained spin $S$ model, we simply have $\alpha=2S+1$. However, for constrained systems we have $\alpha_S<2S+1$. For example, in the PXP model with $S=1/2$, it was shown that $\alpha=\phi<2$, where $\phi$ is the golden ratio, since the Hilbert space dimension scales according to the Fibonacci or Lucas numbers \cite{Turner2018}. 
It can be shown analytically that the quantum dimension for both the QLM and TSM is given by
\begin{align}\label{eq:QD_TSM}
    \alpha_S=2\cos\bigg(\frac{\pi}{4S+3}\bigg),
\end{align}
(see Appendix~\ref{app:QD} for details) which  converges to $2$ as $S\to\infty$; see Table~\ref{tab:QD} for examples. This has a clear physical cause: if the value of the leftmost site is $m$, we can only glue to it a site with spin value $-m$ or $-m-1$ without violating Gauss's law. For a finite $S$, we encounter a further limitation when $m=S$, as $-S-1$ is not a possible spin eigenvalue. However, for infinite $S$ this is not an issue and there are always two different ways of adding a new site. Hence, going from $L$ to $L+1$ doubles the number of states and the quantum dimension is $2$.

For the generalized PXP model~\eqref{eq:PXP}, an analytical expression can be obtained for the quantum dimension as $\alpha_S=\big(1+\sqrt{1+8S}\big)/2$ (see Ref.~\cite{wenwei18TDVPscar} and Appendix~\ref{app:QD} for details). In contrast to the case of the TSM/QLM, the Hilbert space dimension of the spin-$S$ PXP model becomes infinite as $S\to \infty$ and can be well approximated as $\sqrt{2S}$, which is close to the square root of the expected result of $2S+1$ for the unconstrained spin $S$. Again, the physical interpretation is straightforward. In that limit, most of the Hilbert space is taken up by states of the form $\ket{m_1, -S,m_3,-S,m_4,\ldots}$ and $\ket{-S,m_2, -S,m_4,-S,\ldots}$, where  the $m_i$ can take any value between $-S$ and $+S$. The number of these states scales as $2(2S+1)^{L/2}-1$, suggesting a quantum dimension of $\sqrt{2S+1}$. In the large $S$ limit, where this approximation is meaningful, we recover $\alpha_S\approx \sqrt{2S}$ in agreement with the exact expression.

\begin{table}[H]
\center
\begin{tabular}{|l|c|c|c|c|c|c|}
\hline
$S$ & 1/2 & 1 & 3/2 & 2 & 5/2 & 3 \\
\hline
\hline
Unconstrained & 2 & 3 & 4& 5& 6& 7 \\
\hline
QLM/TSM & 1.6180 & 1.8019 & 1.8794& 1.9190& 1.9419& 1.9563 \\
\hline
PXP & 1.6180 & 2 & 2.3028 & 2.5616 & 2.7913 & 3 \\
\hline
\end{tabular}
\caption{Quantum dimension for the various constrained spin models investigated in this work. For the QLM and TSM, the quantum dimension converges towards $2$ for $S\rightarrow \infty$, while for the PXP and the unconstrained models it is unbounded.} \label{tab:QD}
\end{table}

\begin{figure}[t!]
\centering
\includegraphics[width=0.40\linewidth]{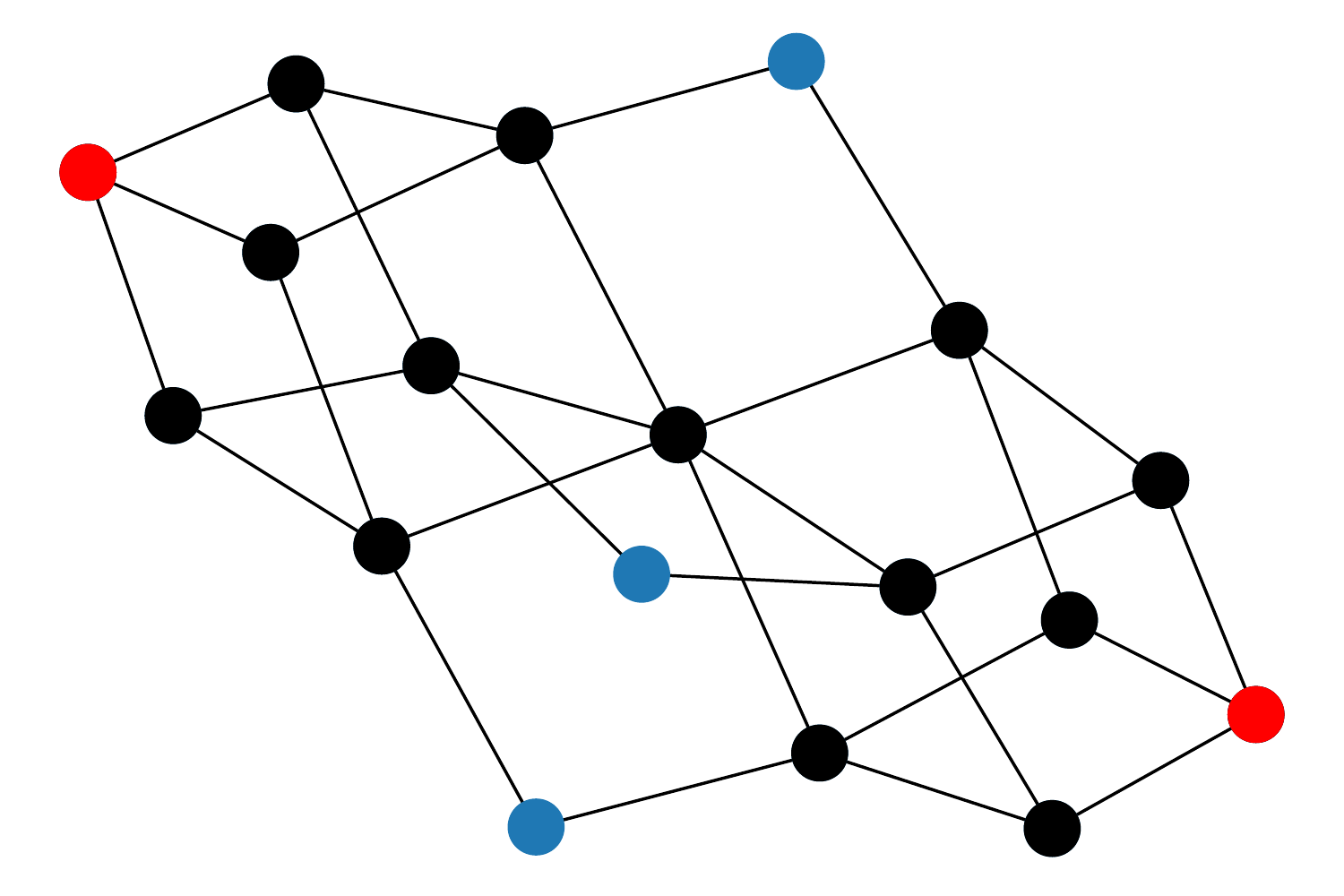}
\includegraphics[width=0.48\linewidth]{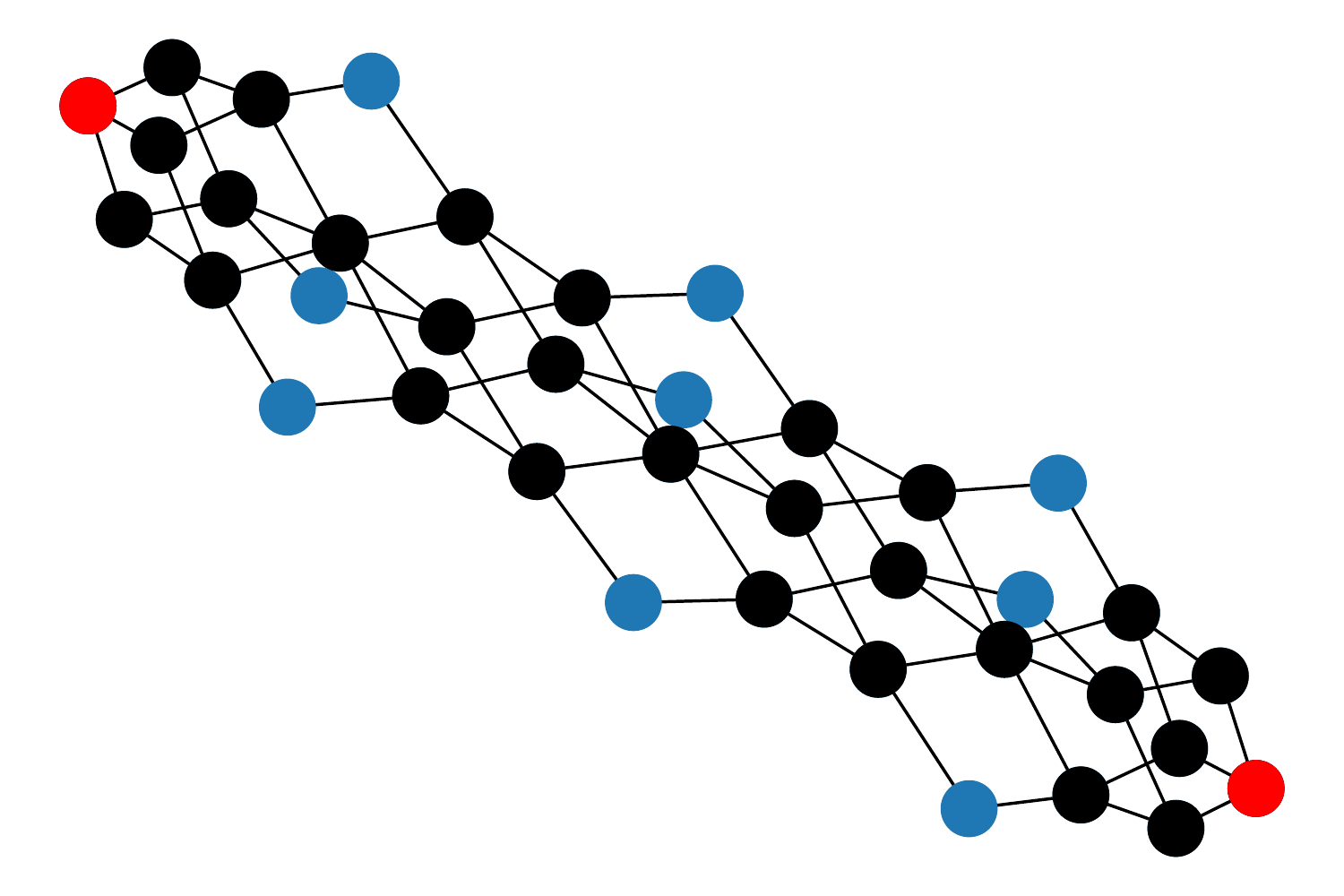}\\
\includegraphics[width=0.6\linewidth]{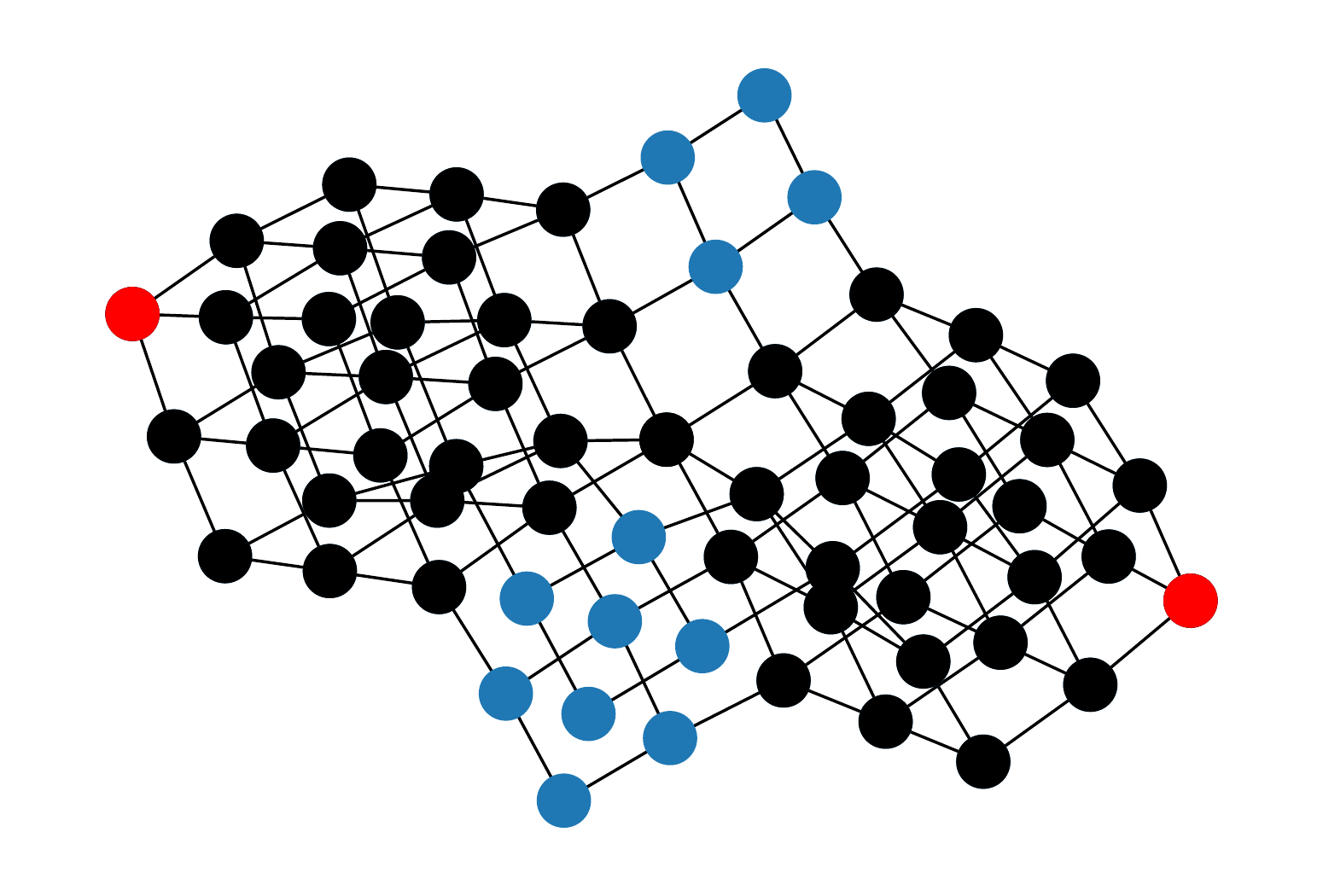}
\caption{Graph of the PXP model with $S=1/2$ (top left), of the QLM/TSM with $S=1$ (top right) and of the PXP model with $S=1$ (bottom), all for $L=6$. The black vertices show the largest hypercubes/hypergrids in the graphs. The red vertices show the best reviving states.}
\label{fig:PXP_LGT_graph}
\end{figure}

This difference highlights that the QLM and TSM are fundamentally different from the generalized PXP model, especially as $S$ becomes large. This difference in the Hilbert space structure can be investigated further by looking at the adjacency graph of the Hamiltonian. In the spin-$1/2$ PXP model, the graph structure consists of two hypercubes of dimension $L/2$ joined at a single vertex (the state $\ket{{\rm CP}}=\ket{-1/2,-1/2,\ldots,-1/2}$), with the rest of the Hilbert space acting as bridges between these two cubes (see Fig.~\ref{fig:PXP_LGT_graph}). At the opposite corner of each cube we find one of the two extreme vacua $\ket{0_-}$ (usually called N\' eel states for that model). A single hypercube holds perfect revivals on its own as long as all the hopping strengths are identical, and two stitched hypercubes also have finite revivals in the thermodynamic limit. As such, it has been conjectured that the revivals in the spin-$1/2$ PXP model are due to its proximity to this toy model of two joined hypercubes \cite{desaules2021hypergrid}. The dynamics can then be thought of as state transfer from the N\'eel state to the shared vertex in the first hypercube, and then from it to the other N\'eel state and back.
For the generalised PXP model in Eq.~(\ref{eq:PXP}), as $S$ is increased these two hypercubes turn into two hypergrids of order $2S+1$ and dimension $L/2$, so equivalent to the graph of a free paramagnet with spin $S$ and $L/2$ sites (see Fig.~\ref{fig:PXP_LGT_graph}). The state shared between the two hypergrids is always $\ket{-S,-S,\ldots,-S}$, and the opposite corners in each hypergrid are the same as the extreme vacua: $\ket{S,-S,\ldots,S,-S}$ and $\ket{-S,S,\ldots,-S,S}$. These two states are also the ones displaying revivals in that model, and the picture of consecutive state transfer in each hypergrid still holds.

In contrast, for the spin-$S$ QLM and TSM, as $S$ is increased we still have hypercubes of dimension $L/2$, but their number increases. Indeed, instead of two hypercubes there are $4S$ of them in a line pattern (see Fig.~\ref{fig:PXP_LGT_graph}). Each hypercube shares states with two neighbors, except the two hypercubes at the end of the chain. The ``unpaired'' states at the corners of these cubes are then the extreme vacua. The state located at the middle of the chain is always nondegenerate and corresponds to either the physical vacuum (for integer spin) or the physical charge-proliferated state (for half-integer spin). In any case, all the vacua and charge-proliferated states are always located at the intersection of two hypercubes. 
For these models the simple picture of dynamics being consecutive state transfer along the chain of hypercubes works as well. It also explains why we get revivals in the TSM despite the unconstrained model showing none for higher $S$. Due to the constraint, the relevant dynamics happens in the hypercubes, which mimic effective systems with spin-$1/2$. In that case, the unconstrained TSM also has perfect revivals.

\begin{figure}[t!]
\centering
\includegraphics[width=\linewidth]{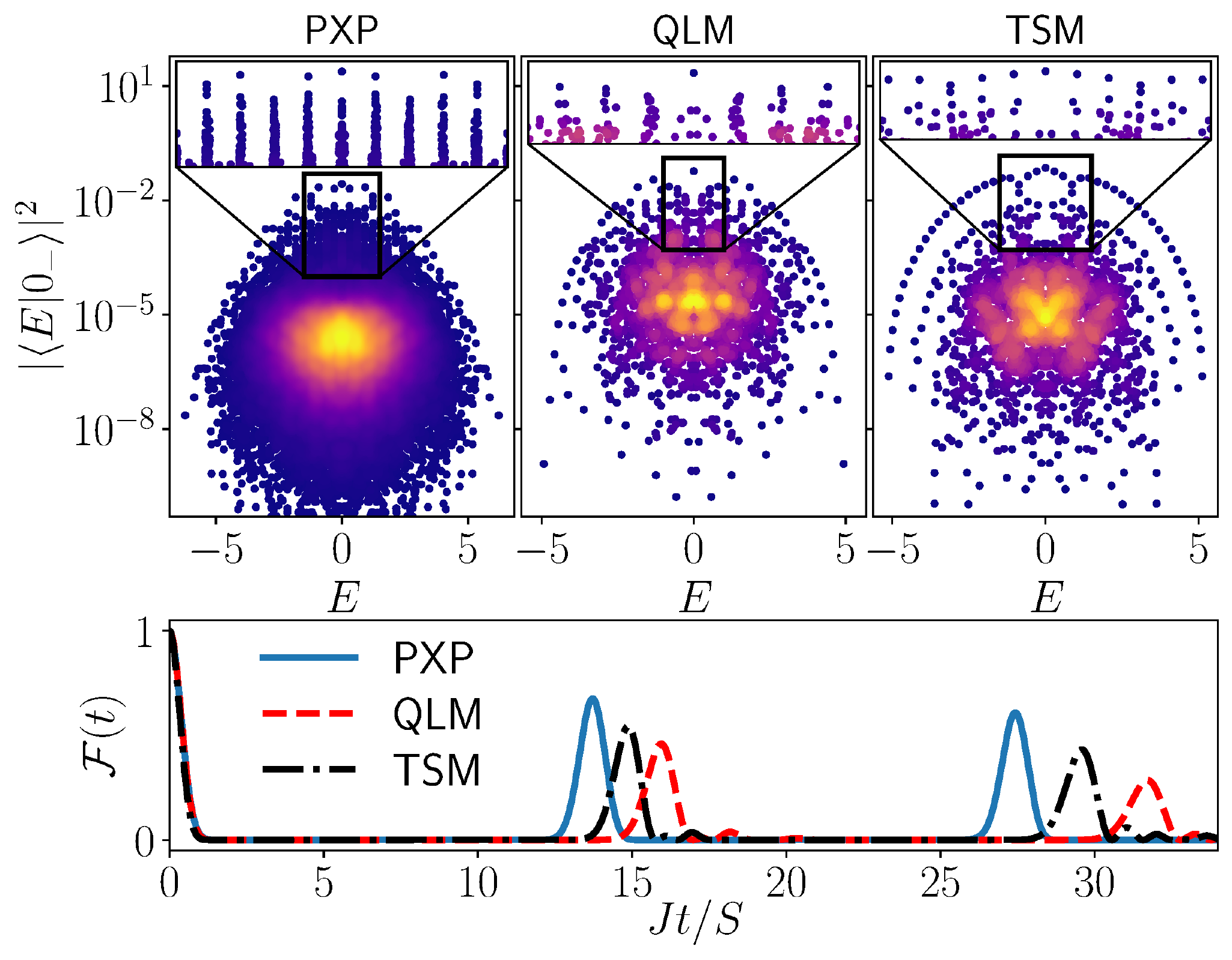}
\caption{Comparison of the PXP, QLM, and TSM for $L=16$ and $S=3/2$. For the PXP model, we set $\Omega=1/\sqrt{S(S+1)}$ to match the QLM. Even if all three models exhibit $2SL+1=49$ towers of states, their characteristics differ greatly. These differences translate to the revivals from the $\ket{0_-}$ state.}
\label{fig:PXP_LGT}
\end{figure}

This difference in the graph structure further amplifies the dissimilarities between the PXP model on one hand, and the QLM and TSM on the other. It also implies that the physical interpretation of the relevant classical limit is different for PXP than for the QLM and TSM, as we will show below. Nevertheless, there are still some striking similarities between them. The main one is the number of special eigenstates (or towers of states), which is equal to $2SL+1$ for all of them. This number simply corresponds to the distance in the graph between the two extreme vacua plus one. In all models, these special eigenstates can be well-approximated by the forward scattering approximation from Ref.~\cite{Turner2018}. As shown in Appendix~\ref{app:S1_detailed}, the revivals can be further enhanced by adding a perturbation whose form is inspired by the forward scattering, as previously done in Refs.~\cite{Khemani2018, Choi2018}. 
However, even if the number of towers of states is the same, there are some significant differences in their structure, as we illustrate in Fig.~\ref{fig:PXP_LGT}. Indeed, for the PXP model theses towers are dense, but very narrow in energy and extending far above the bulk of the spectrum in terms of overlap with the initial state. For the QLM, this is not the case, as the spread in energy among the states in the same tower is much larger. This leads to a much quicker dephasing and decay of the fidelity revivals. Finally, the TSM shows a picture closer to the one of the PXP model with $S=1/2$: the towers of states are relatively sparse, with a single eigenstate at the top which is well-separated from the rest.

\subsection{Classical limit} 

The graph representation in Fig.~\ref{fig:PXP_LGT_graph} is helpful for constructing the classical limit of scarred dynamics. For the PXP model, it was shown that the classical limit corresponds to a periodic orbit~\cite{wenwei18TDVPscar}. The mapping to a classical dynamical system was obtained using the time dependent variational principle (TDVP) \cite{Haegeman2011}. The Ansatz used in Ref.~\cite{wenwei18TDVPscar} simply corresponds to setting coherent spin states $\ket{(\theta,\phi)}=e^{i\phi S}e^{i\phi \hat{s}^z}e^{-i\theta \hat{s}^x}\ket{-S}$ on odd and even sites separately and projecting that wave function into the constrained Hilbert space.
The Ansatz incorporates the constraint and it was shown that it yields a compact representation in terms of a matrix product state with the bond dimension equal to $2$, regardless of the magnitude of the spin. The intuitive picture behind the Ansatz is that each sublattice acts as a big spin, whose evolution is dependent on the value of the spin corresponding to the other sublattice~\cite{TurnerSym}. These spin coherent states naturally describe each hypergid on its own (which corresponds to having excitations on only one sublattice), while the constraint incorporates the coupling between them.

For the QLM and TSM, however, this two-angle description is not enough to capture the dynamics. Indeed, as hinted by the graph structure, we do not have two big spins $SL$ coupled together, but instead $4S$ spins-$L/2$. Hence, in order to properly describe the evolution of this system, we conjecture that we would have to keep track of $8S$ angles, i.e., two per big spin, or alternatively $4S$ per sublattice.
While this already makes the definition of a classical limit that could expand all the way to the Schwinger model basically impossible, the bond dimension needed will also be problematic. 
Indeed, unlike for the generalized PXP model, there is no simple way to encode the constraint into a fixed bond dimension MPO for any $S$. In Appendix~\ref{app:MPO}, we present an argument that for the TSM and QLM, the minimum bond dimension required to encode the constraint grows linearly with $S$. This shines light on the profound difference of the constraint between the PXP model and the QLM, and shows that obtaining a classical limit using a TDVP Ansatz can quickly become intractable as $S$ increases. 

As such, developing a classical limit using TDVP for $S>1$ for the QLM and TSM appears to be a nontrivial task due to the form of the constraint. However, we can still use a ``mean-field-like'' approximation that confines the dynamics to the span of the desired TDVP manifold. This was done for the PXP model with $S=1/2$ using the symmetric subspace approximation \cite{TurnerSym}. While in the TDVP Ansatz each sublattice is characterised by an angle that sets the probability of having $m=+1/2$ on any site, in the quantum versions we simply create a basis where each state is a symmetric superposition of all states having a set number $n_i$ of site $m=+1/2$ on each sublattice. Each state in the basis is then characterized by the pair $(n_1,n_2)$, and we obtain a basis of the span of the TDVP manifold. It was additionally shown that the dynamics in that symmetric subspace $\mathcal{K}$ corresponds to requantizing the semiclassical TDVP dynamics \cite{TurnerSym}.  

\begin{figure}[t!]
\centering
\includegraphics[width=\linewidth]{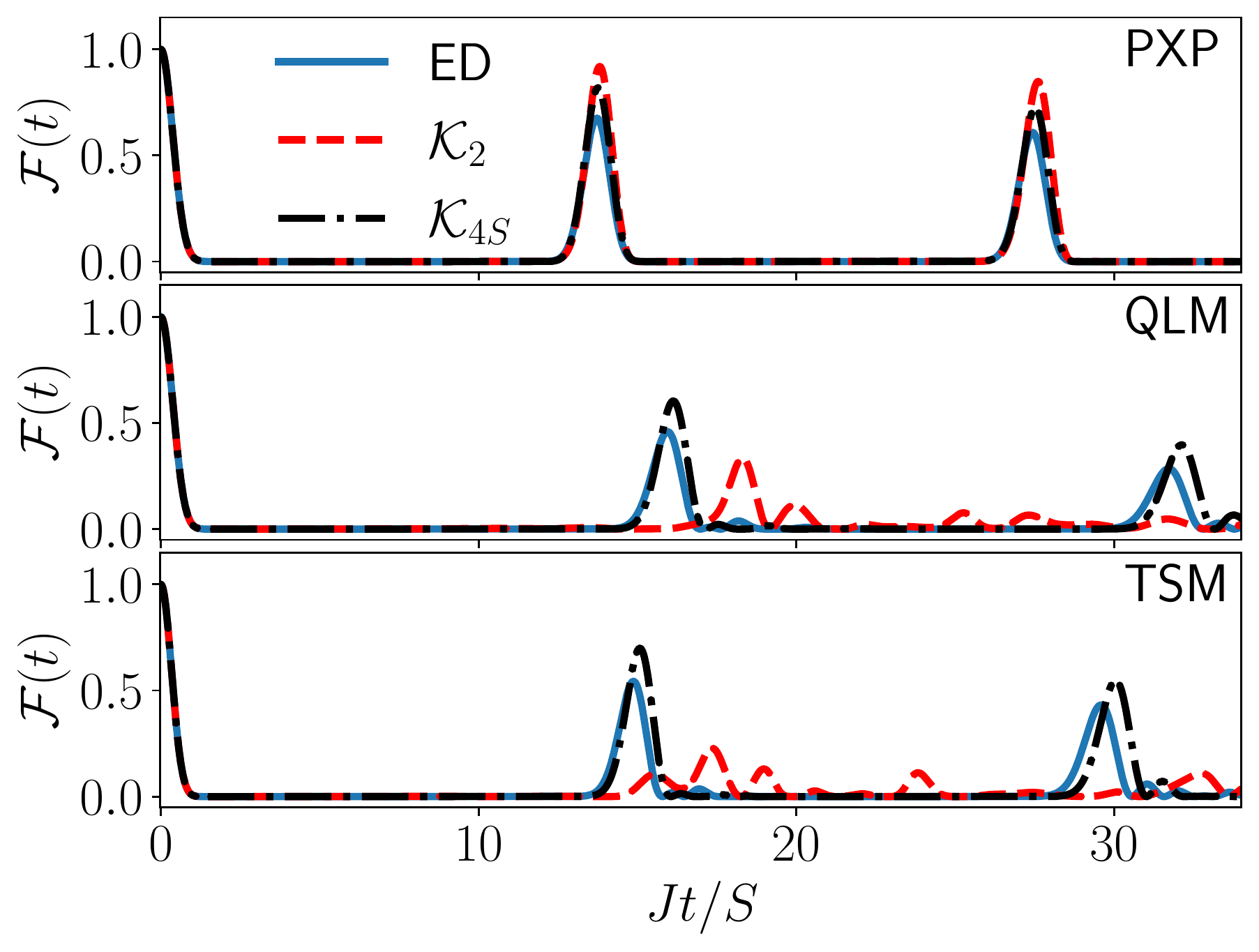}
\caption{Dynamics of the fidelity after a zero-mass zero-$g$ quench from the $\ket{0_-}$ state in the PXP, QLM, and TSM with $L=16$ and $S=3/2$. The solid line corresponds to data from exact diagonalization, while the dashed and dashed-dotted lines correspond to the dynamics in the two different symmetric subspaces. The smaller subspace $\mathcal{K}_2$ already captures the behavior of the PXP model quite well but fails to do so for the QLM and TSM. Meanwhile, the larger subspace $\mathcal{K}_{4S}$ gives good results for all three models.}
\label{fig:perm_fid}
\end{figure}

Here, we expand this approximation to the PXP model with higher $S$ by using coherent spin states, as is the case in the corresponding TDVP Ansatz of Ref.~\cite{wenwei18TDVPscar}. We can then form a basis of these constrained coherent states, where each basis state is characterized by two numbers. These are directly related to the $\theta$ angles in the TDVP Ansatz (see Appendix~\ref{app:sym_subs}).  
We now obtain the subspace $\mathcal{K}_2$ spanned by these basis states. We can then project the Hamiltonian into this subspace and compare the evolution within it with the one in the full Hilbert space. It gives a very good description of the dynamics from the $\ket{0_-}$ state, as can be seen in Fig.~\ref{fig:perm_fid}. We emphasize that this is highly nontrivial as the projection into the subspace corresponds to a reduction of the effective dimension of the Hilbert space from $40477$ states (taking into account only the relevant translation and reflection sectors) to $238$ states. On the other hand, this completely fails to capture the dynamics in both the TSM and QLM models, even though the initial Hilbert space dimension is much smaller with only $1866$ (symmetry resolved) states. In order to get a decent approximation of the dynamics in these models, we have to expand our approximation to keep more information. 

\begin{figure}[t!]
\centering
\includegraphics[width=\linewidth]{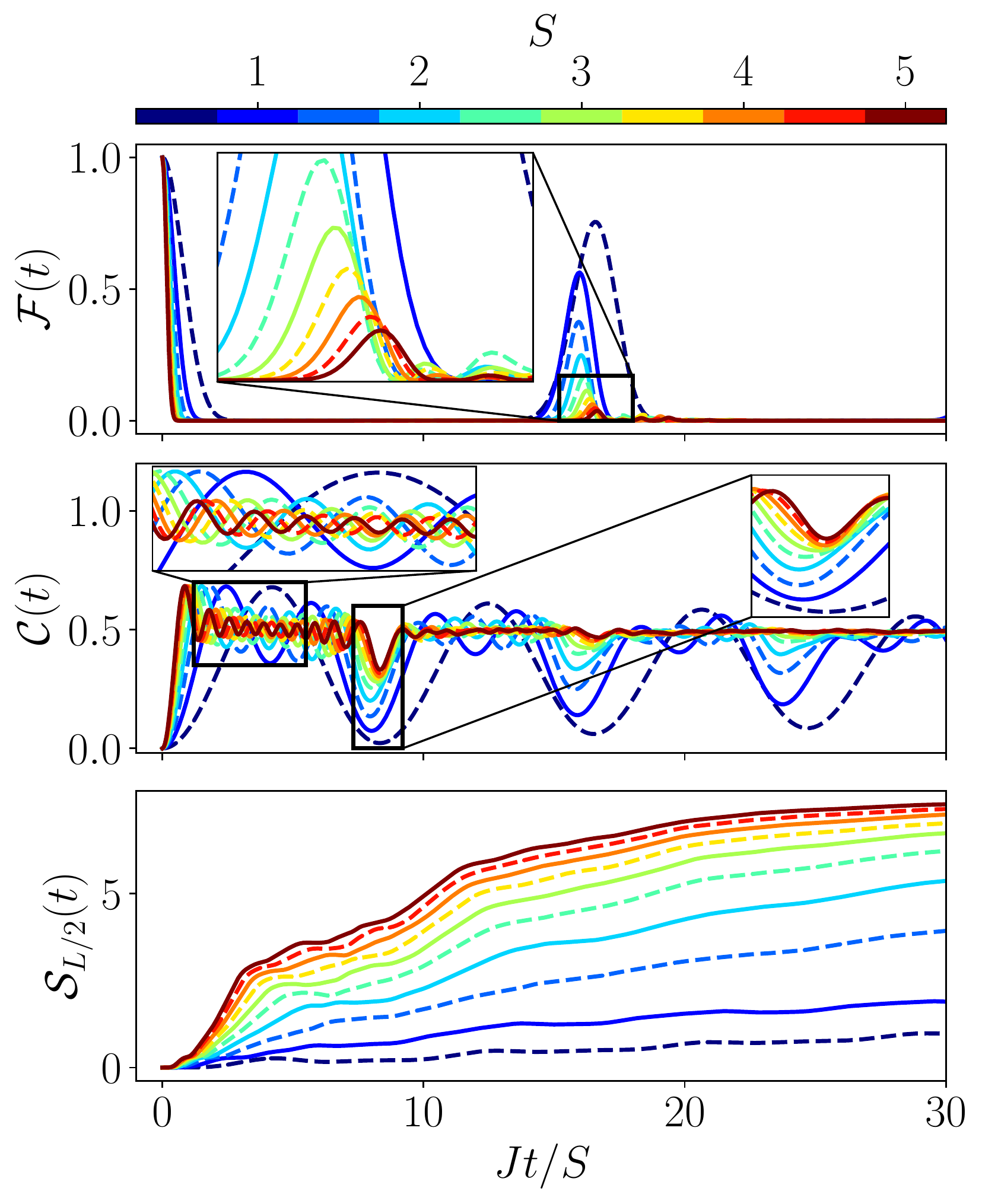}
\caption{Dynamical properties for zero-mass zero-$g$ quenches from the extreme vacuum $\ket{0_-}$ for $L=20$ and various values of $S$ in the QLM. The solid lines show integer $S$, while the dashed lines denote half-integer $S$. No clear distinction can be made between these two cases, with both the fidelity revival amplitude and the difference of the chiral condensate from the equilibrium value decreasing monotonically with $S$.}
\label{fig:LGT_S_N}
\end{figure}
\begin{figure}[t!]
\centering
\includegraphics[width=\linewidth]{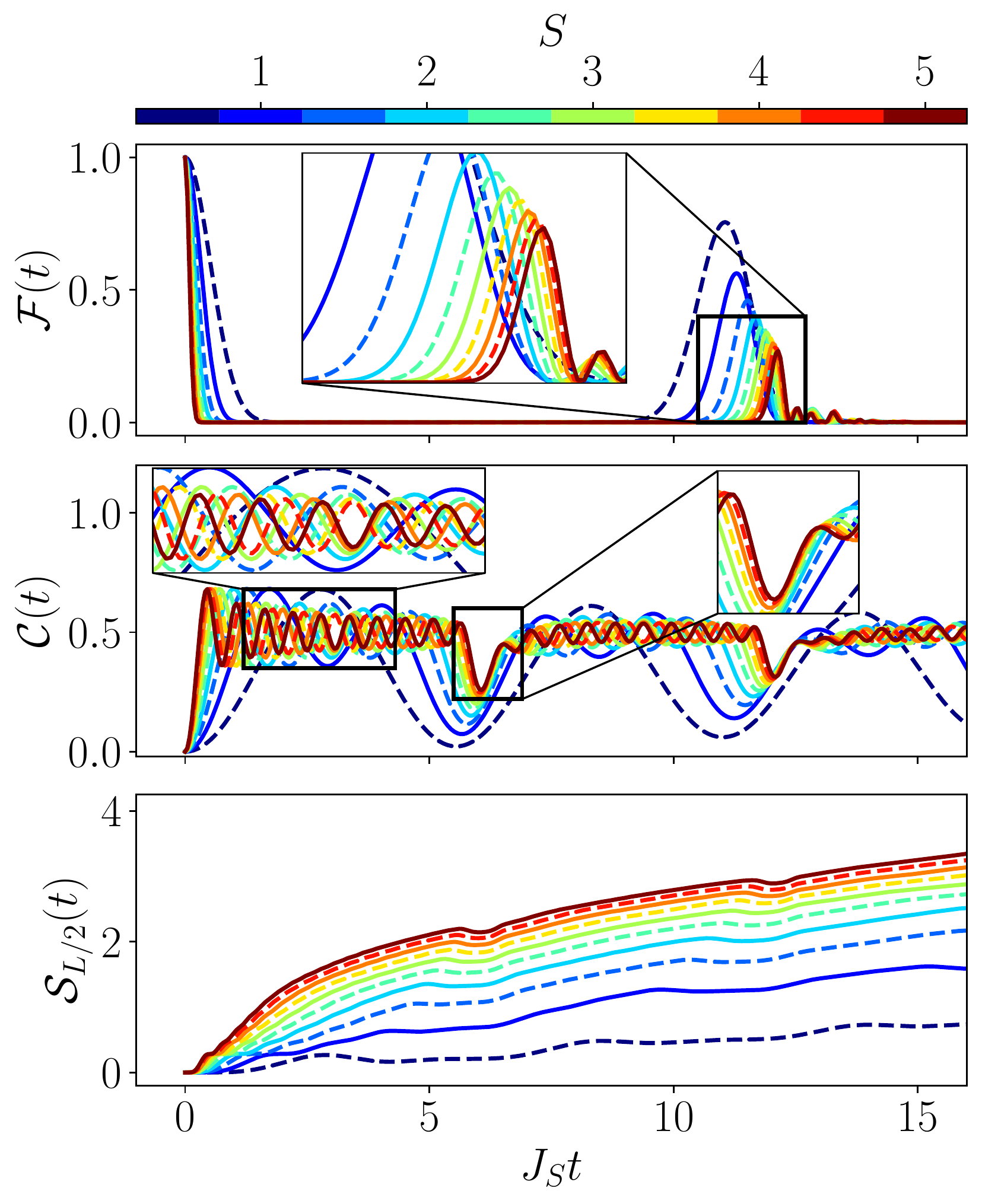}
\caption{Dynamical properties for quenches from the extreme vacuum $\ket{0_-}$ for $L=20$ and various values of $S$ in the TSM, with $J_S=J/\sqrt{S(S+1)}$. The solid lines show integer $S$, while the dashed lines denote half-integer $S$. The results are qualitatively the same as in the QLM, however all three quantities indicate much stronger scarring in the TSM.
}
\label{fig:LGT_eq_N}
\end{figure}

As mentioned previously, in the TDVP Ansatz for these models we want to capture the dynamics of $4S$ effective spins-$L/2$ corresponding to the hypercubes in the graph. In order to do this, we devise a larger subspace in which we keep track of the number of each spin eigenvalue $m=-S$ to $m=S$ for each sublattice (see Appendix~\ref{app:sym_subs}). We end up with a total of $4S$ numbers characterizing each basis state of our subspace. This is exactly half of the number of angles in the conjectured TDVP Ansatz, as we do not need to incorporate the phase angles $\phi$ into our basis since they can evolve freely during the quantum evolution.
The results of the dynamics projected into this new subspace $\mathcal{K}_{4S}$ can now be compared with the true dynamics in the full Hilbert space (see Fig.~\ref{fig:perm_fid}). We see that the behavior much more closely matches the actual fidelity for the TSM and QLM. This also gives a good result for the PXP model, however this is expected as $\mathcal{K}_2\subset \mathcal{K}_{4S}$.

This demonstrates that a low-dimensional subspace that captures the relevant dynamics from a zero-mass zero-$g$ quench can be devised simply by studying the adjacency graph structure of the Hamiltonian. This structure could also theoretically be used to obtain a classical limit for these constrained models using the corresponding TDVP Ansatz.
These constructions then show that for the PXP model, the relevant classical limit is much simpler than for the QLM and TSM.

As such, we have demonstrated in this section that the spin-$S$ PXP model is fundamentally different from its TSM and QLM counterparts. Nevertheless, we can still utilize tools developed for the PXP model in order to better understand the TSM and QLM, and to even enhance scarring behavior in them. For further details, see Appendix~\ref{app:S1_detailed} for results on the forward scattering approximation and enhanced scarring through algebra-correcting perturbations in the case of spin-$1$ QLM and TSM (recall that for $S=1$, the QLM and TSM are identical).

\section{Kogut--Susskind limit ($S\to\infty$)}\label{sec:S_lim}

In Sec.~\ref{sec:resonant}, we have shown that the extreme vacuum shows persistent revivals in both the TSM and QLM for various values of $S$. Here we study the scaling of these revivals as $S$ is varied. Figures~\ref{fig:LGT_S_N} and \ref{fig:LGT_eq_N} show how various quantities change with $S$ when the system size is fixed to $L=20$. In both models the scarring signature gets weaker as $S$ increases. The revivals in the fidelity and chiral condensate get worse with $S$, while the entropy growth also becomes faster. This is not surprising, since the fraction of the scarred subspace within the total Hilbert space decreases with $S$. Comparing these quantities between the two models also shows clearer scarring in the TSM.

These results also show an interesting behavior in the dynamics of the chiral condensate for both the TSM and QLM. In between the main oscillations at twice the revival frequency, we also see oscillations with smaller amplitudes in the chiral condensate. These can be explained by considering the quantum evolution as the wave function propagating along the hypercubes forming the ``backbone'' of the graph (see Sec.~\ref{sec:comp_PXP} for details). Due to these hypercubes, we can see the dynamics as resembling state transfer in each one of them sequentially before finally reaching the other side of the graph. This is of course a very crude description as it ignores large parts of the Hilbert space. However, it is useful in helping us understand the intermediate states between the extreme vacua. As all states have a two-site periodicity, we will describe them by the value on the first two sites, i.e., $\ket{M_1,M_2}=\ket{M_1,M_2,M_1,M_2,\ldots,M_1,M_2}$. Starting from $\ket{0_-}=\ket{S,-S}$, we have (approximate) state transfer to $\ket{S-1,-S}$, then to $\ket{S-1,1-S}$, then $\ket{S-2,1-S}$, and so on until we reach the other end of the chain with $\ket{0_-^\prime}=\ket{-S,S}$. These intermediate states with periodicity two alternate between vacua and charge-proliferated states, leading to the oscillations in the chiral condensates. As the wave function is not exactly on these states but spread amongst states in the same ``slice'' of the graph (i.e., at the same distance of the two extreme vacua), we get less extreme values of the chiral condensate. Nonetheless, this sequential state transfer picture allows us to predict the number of these low-amplitude oscillations. As there are $4S-1$ states between the extreme vacua, we expect to see the same number of local extrema in $\mathcal{C}(t)$.  Out of these, $2S$ should be maxima and $2S-1$ should be minima. This perfectly describes the dynamics of the chiral condensate and shows that this simple picture of propagation along the backbone of the graph is a good approximation of the dynamics from the extreme vacua.

\begin{figure}[t!]
\centering
\includegraphics[width=\linewidth]{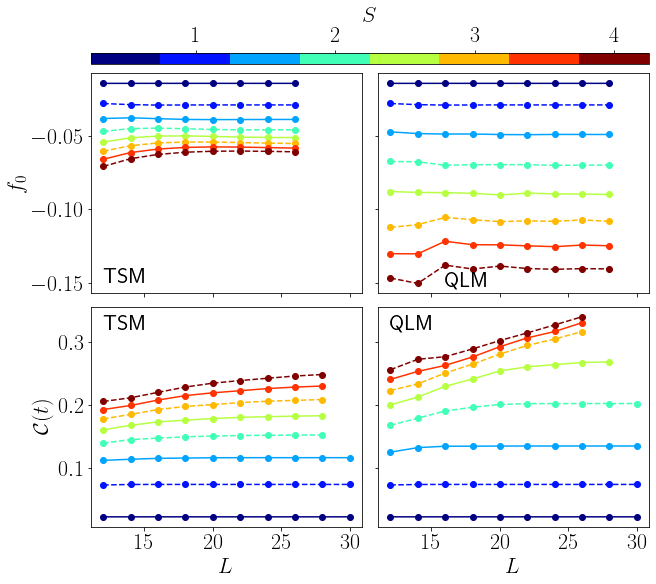}
\caption{Fidelity density $f_0=\ln(\mathcal{F}_0)/L$ and chiral condensate values at their first respective peak after a quench from the extreme vacuum for various values of $S$ and $L$. The solid lines show integer $S$, while the dashed lines denote half-integer $S$. Both quantities eventually converge with $L$, allowing us to make predictions about the behavior in infinite systems.}
\label{fig:LGT_conv_all}
\end{figure}
\begin{figure}[t!]
\centering
\includegraphics[width=\linewidth]{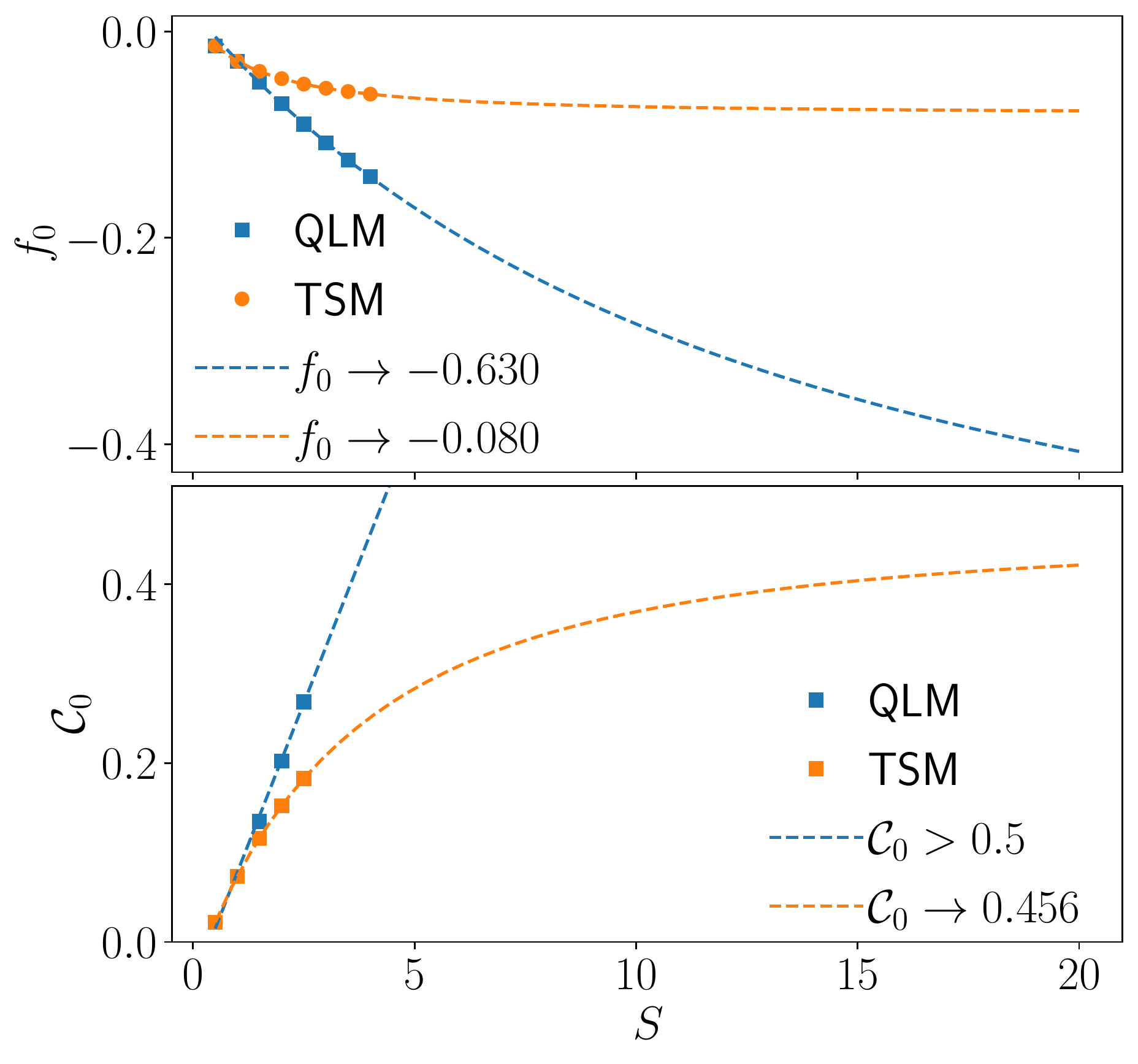}
\caption{Extrapolation of the behavior of the fidelity density and chiral condensate value at their first respective revivals. The data is fitted with the function $a+b/(c+x)^\gamma$, with $\gamma=3/2$ for $f_0$ and $\gamma=2$ for $\mathcal{C}$. Changing $\gamma$ between 1 and 3 does not lead to significant changes for the infinite $S$ value. These results suggest that the QLM becomes fully ergodic for infinite $S$, while the truncated Schwinger model still shows scarring in that limit.}
\label{fig:LGT_conv_FM}
\end{figure}

While the scarring behavior getting weaker is expected, the real question is whether it disappears as $S\to \infty$ or if the two models will still display ergodicity breaking in that limit. Let us first get rid of finite-size effects and then scale $S$. In order to do this, we investigate quantities that converge towards a finite value as $L\to \infty$ and then study the scaling of the converged values.
For that we can use the fidelity density, defined as $f_0=\ln(\mathcal{F}_0)/L$ with $\mathcal{F}_0$ the first fidelity-revival amplitude. We can also use the value of the chiral condensate at its ``revival'' (which happens after a half revival period).
For a fully ergodic dynamics we would expect fast thermalization to a state that reproduces the predictions of thermal ensembles at infinite temperature, giving $\mathcal{C}_0=0.5$ for all $S$. The wave function should also be spread evenly in the Hilbert space, giving a fidelity of $1/\mathcal{D}=\alpha^{-L}$, with $\mathcal{D}$ the Hilbert space dimension and $\alpha$ the quantum dimension. This results in a fidelity density equal to $f_0=\ln(\alpha^{-L})/L=-\ln(\alpha)$. As shown in Sec.~\ref{sec:comp_PXP}, we know that $\lim_{S\to\infty}\alpha=2$, thus the ergodic value is $f_0=-\ln(2)\approx -0.6931$.

Figure~\ref{fig:LGT_conv_all} shows the convergence of these two quantities with $L$ for different values of $S$ in both the TSM and QLM. For the fidelity density the convergence in $L$ is quite fast, and we can use the data for $S=1/2$ to $S=4$. However, for $\mathcal{C}_0$ the convergence is much slower, and we can only extrapolate the value at infinite $L$ for $S$ up to $5/2$.

For the TSM, both quantities are relatively well-approximated by a power-law scaling of  $a+b/(c+x)^\gamma$ with $1<\gamma<3$; see Fig.~\ref{fig:LGT_conv_FM}. While we are not aware of any predicted scaling that would fix $\gamma$, we note that the extrapolated value at $S\to\infty$ varies relatively little with $\gamma$.  We find that at $S\to \infty$, the absolute value of the fidelity density is smaller than that of the predicted value by an order of magnitude, indicating ergodicity breaking. While the chiral condensate value also seems to indicate a lack of thermalization, the small number of data points and proximity to the thermal value make it impossible to draw any strong conclusions.

Nonetheless, the contrast between the two models is stark. Indeed, for the fidelity density the predicted value for the QLM is very close to the ergodic value, well within the error range of the extrapolation. As for $\mathcal{C}_0$, the data available shows a rapid increase towards the thermal value. As there is no evidence or physical reason as to why the value should rise above the thermal one while approaching it from below, this suggests that this behavior should change for some higher value of $S$. 
Overall, the results of the finite-$S$ scaling for the QLM are relatively imprecise, but they seem to suggest that any dynamical signature of scarring may vanish as $S$ goes to infinity. In contrast, in the case of the TSM our results strongly suggest that scarring persists in the limit $S\to\infty$.

This highlights one of the limitations of our graph theoretical approach. As it does not directly take into account the strength of the matrix elements nor their structure outside of the dominant hypercubes, it falls short at predicting different behaviors for the TSM and QLM as $S$ increases. 

\section{Conclusions and outlook}\label{sec:conclusion}
In this work, we have investigated the spin-$S$ QLM and TSM, two different constrained spin models based on the lattice Schwinger model. In both models, we have shown clear signatures of quantum many-body scarring when quenching from the extreme vacua to zero mass and gauge coupling strength. We have also shown that both the physical vacua and the charge-proliferated states can exhibit detuned scarring for finite values of the mass and gauge coupling. Both resonant and detuned scarring have been previously observed in the spin-$1/2$ PXP model \cite{Bernien2017,Turner2018,Su2022}, which is recovered from both the TSM and QLM for $S=1/2$. However, for any other value of $S$, these models are different from the generalized spin-$S$ PXP model previously studied in the literature \cite{wenwei18TDVPscar}.
This difference can be made evident by investigating the structure of the adjacency graph of the corresponding Hamiltonians. Using this approach, we have also proposed a procedure to construct a classical limit for the QLM and TSM using TDVP, which again differs from the one used for the PXP model. While for the TSM and QLM this classical limit is more complicated to implement, we have shown that its structure is apt at capturing the resonant scarring through a simpler quantum approximation.

Our simulations have also shown that for any $S$ investigated, the TSM exhibits stronger scarring than the QLM in all metrics used. These differences get more pronounced as $S$ increases, and based on a finite-$S$ scaling analysis we predict that the TSM still shows signs of weak ergodicity breaking for $S\to \infty$ while this cannot be ascertained for the QLM. Understanding the source of these differences between the two models would be an interesting goal for future works, as it requires the investigation of the graph structures beyond the dominant subgraphs.

Another question raised by our work is the fate of quantum many-body scars in spin models corresponding to higher-dimensional versions of $\mathrm{U}(1)$ lattice gauge theories with dynamical matter. As the coordination number of the lattice increases, the number of configurations allowed by Gauss's law will rapidly get larger. Thus the constraint should get weaker and the graph structure linked to it should also change drastically. As a consequence, it is currently not known if these models also possess scarring behavior, or if the latter is only a feature present in low-dimensional cases.

Given the massive current drive in implementations of lattice gauge theories on synthetic quantum matter setups \cite{Martinez2016,Muschik2017,Klco2018,Keesling2019,Kokail2019,Goerg2019,Schweizer2019,Mil2020,Klco2020,Yang2020,Zhou2021,Mildenberger2022}, our work further provides insight on the observation of scarring in such realizations.

\begin{acknowledgments}
J.C.H.~and J.-Y.D.~are very grateful to Giuliano Giudici for insightful discussions and valuable comments. J.-Y.D.~would also like to thank Andrew Hallam for fruitful discussions. J.C.H.~acknowledges funding from the European Research Council (ERC) under the European Union’s Horizon 2020 research and innovation programm (Grant Agreement no 948141) — ERC Starting Grant SimUcQuam, and by the Deutsche Forschungsgemeinschaft (DFG, German Research Foundation) under Germany's Excellence Strategy -- EXC-2111 -- 390814868. We acknowledge support by EPSRC grants EP/R020612/1 (Z.P.) and EP/R513258/1 (J.-Y.D.). A.H.~acknowledges funding provided by the Institute of Physics Belgrade, through the grant by the Ministry of Education, Science, and Technological Development of the Republic of Serbia. Z.P.~acknowledges support by the Leverhulme Trust Research Leadership Award RL-2019-015. Statement of compliance with EPSRC policy framework on research data: This publication is theoretical
work that does not require supporting research data.
\end{acknowledgments}

\appendix
\begin{figure}[t!]
\centering
\includegraphics[width=\linewidth]{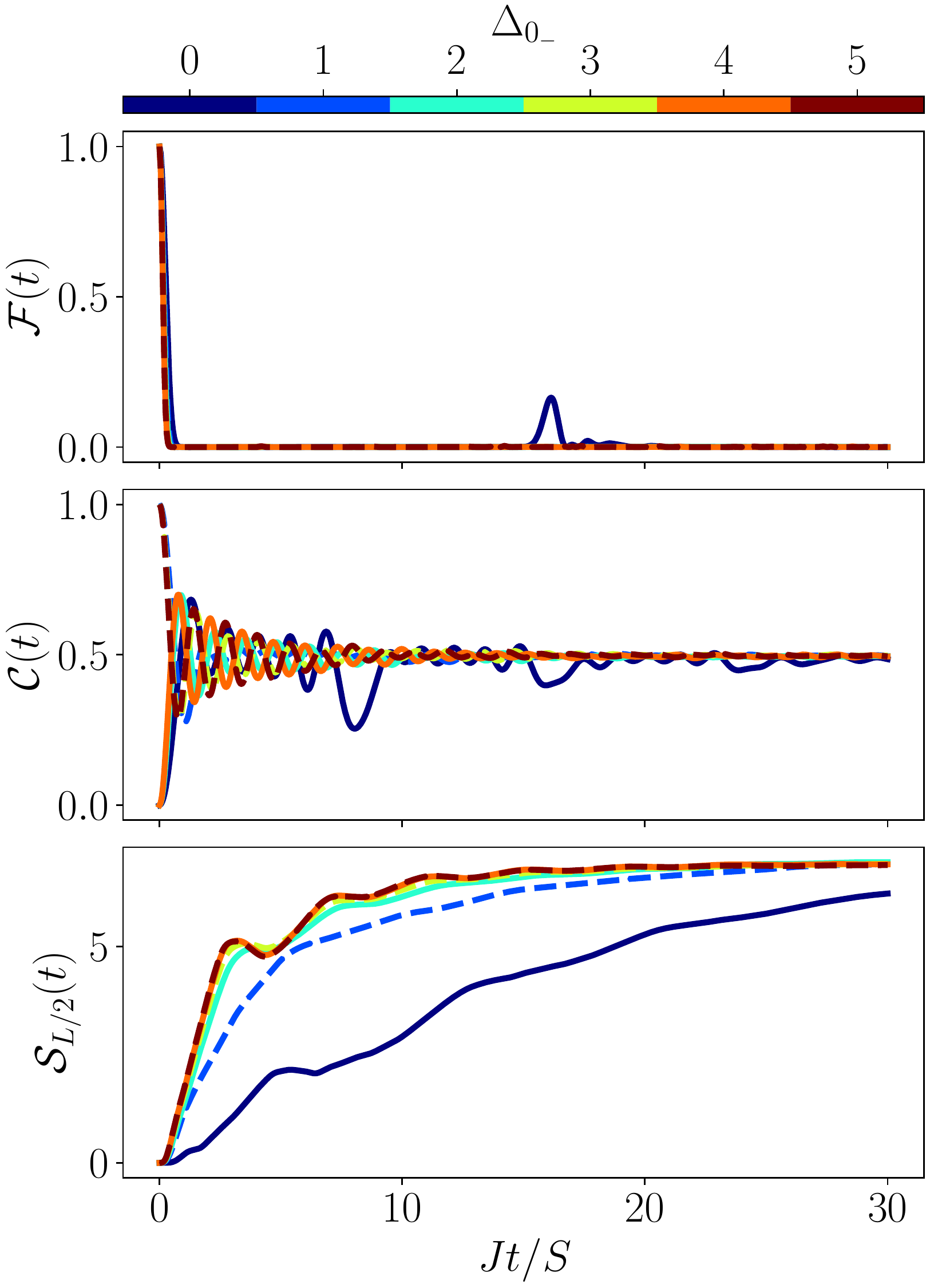}
\caption{Dynamical properties for zero-mass zero-$g$ quenches starting in various vacua (solid line) and charge-proliferated states (dashed lines) for $S=5/2$ and $L=20$ in the QLM. $\Delta_{0_-}$ quantifies their proximity to the extreme vacuum $\ket{0_-}$, the only state showing clear revivals. The entanglement entropy growth also suggests that thermalization happens faster as $\Delta_{0_-}$ increases.}
\label{fig:LGT_S_all_vac}
\end{figure}

\begin{figure}[t!]
\centering
\includegraphics[width=\linewidth]{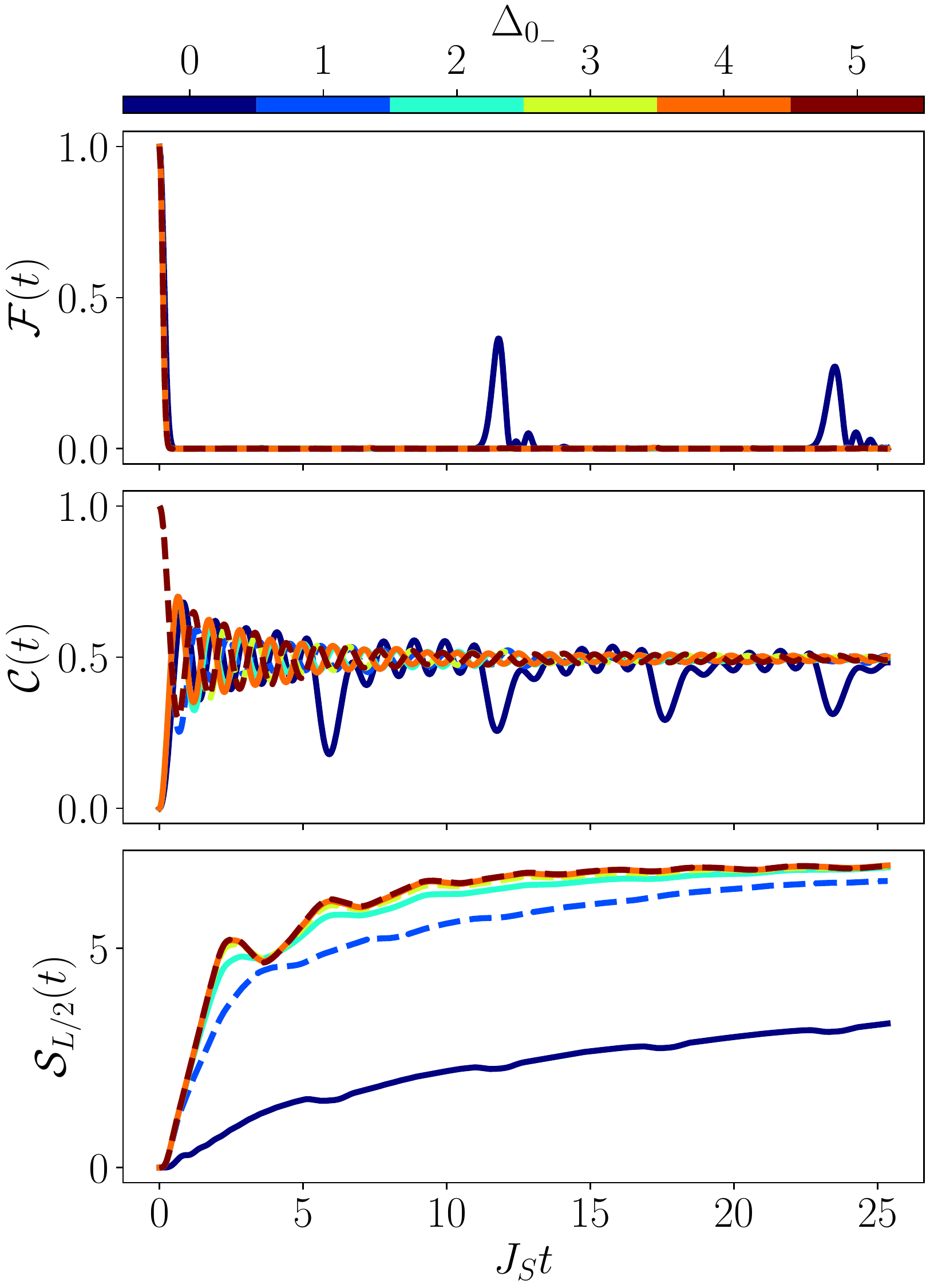}
\caption{Dynamical properties for zero-mass zero-$g$ quenches starting in various various vacua (solid lines) and charge-proliferated states (dashed lines) for $S=5/2$ and $L=20$ in the TSM, with $J_S=J/\sqrt{S(S+1)}$. $\Delta_{0_-}$ quantifies their proximity to the extreme vacuum $\ket{0_-}$, the only state showing clear revivals.  The results are qualitatively very similar to the one in the QLM.}
\label{fig:LGT_eq_all_vac}
\end{figure}

\section{Quenches from other vacua and CP initial states} \label{app:other_states}
In the main text, we have shown the results of zero-mass zero-$g$ quenches from the extreme vacua, the physical vacua, and the charge-proliferated state for the TSM. Similar results for the QLM are available in Ref.~\cite{Desaules2022weak}. For all values of $S$ and $L$ investigated, only the extreme vacua showed signatures of resonant scarring. However, as $S$ is increased there are more and more vacua and charge-proliferated states corresponding to different values of the field. In the spin language, all the vacua have the structure $\ket{M,-M}$, and the charge-proliferated states $\ket{M-1,-M}$. We can quantify these states by their proximity $\Delta_{0_-}$ to the extreme vacuum $\ket{0_-}=\ket{S,-S}$. The extreme vacuum then has $\Delta_{0_-}=0$, the charge-proliferated state with maximum electric field $\ket{S-1,-S}$ has $\Delta_{0_-}=1$, the vacuum state $\ket{S-1,1-S}$ has $\Delta_{0_-}=2$ and so on. Figures~\ref{fig:LGT_S_all_vac} and \ref{fig:LGT_eq_all_vac}  show that the extreme vacuum is the only state showing revivals in the wave function for a zero-mass zero-$g$ quench in both the QLM and TSM. However, the growth of entanglement entropy seems to increase monotonically with $\Delta_{0_-}$, hinting that states in the middle of the graph (far away form the extreme vacua) thermalize faster.

\section{Detuned quenches from the extreme vacua}\label{app:det_0m}

\begin{figure}[t!]
\centering
\includegraphics[width=\linewidth]{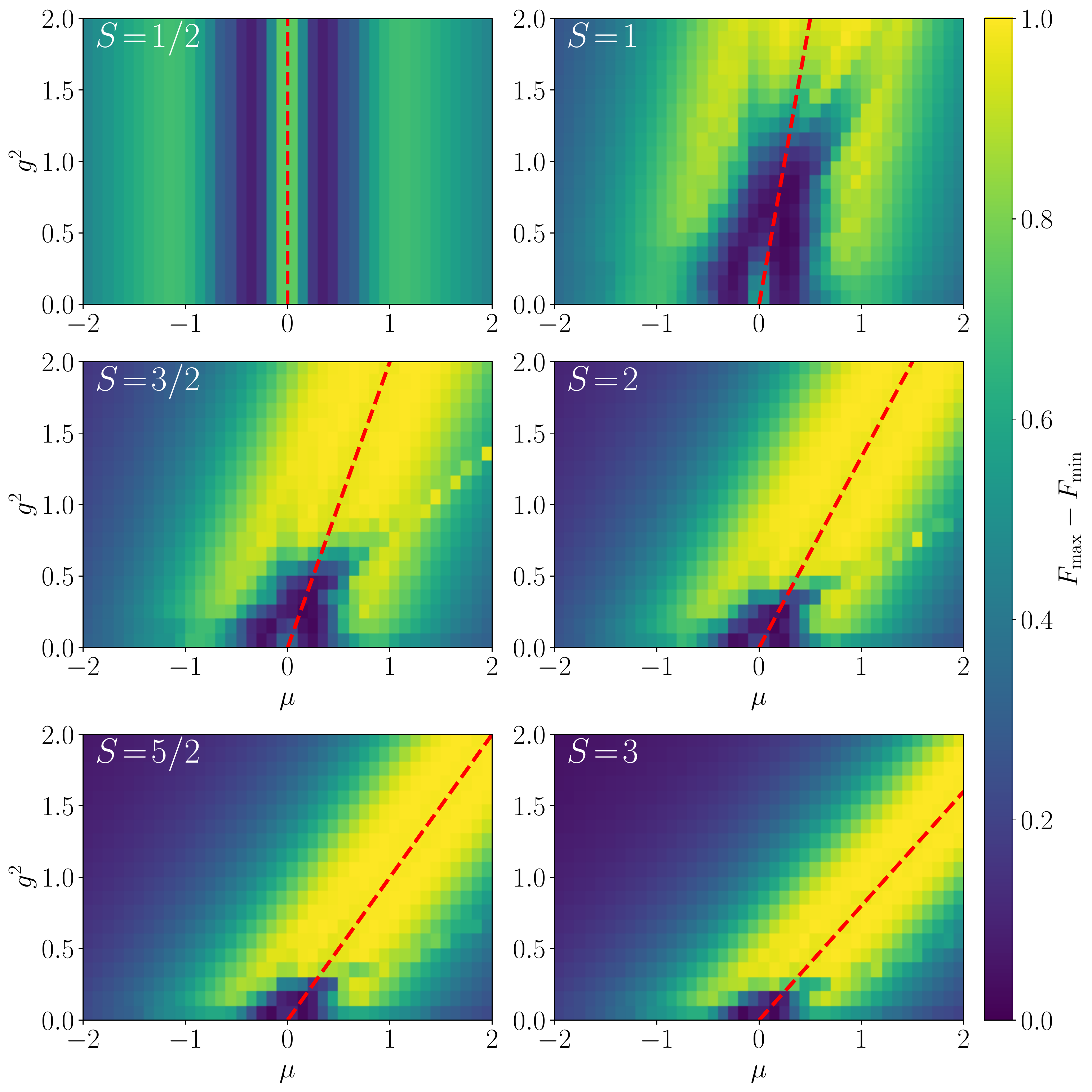}
\caption{Difference between the maximum and minimum amplitudes of the fidelity revival when quenching from the state $\ket{0_-}$ for $L=16$ and various values of $\mu$ and $g^2$ in the QLM. The red dashed line in each plot shows the resonance condition $\mu=(2S-1)g^2/4$ for which the Hilbert space fractures when $\mu,g^2\gg J/\sqrt{2(S+1)}$.}
\label{fig:LGT_quench_0m}
\end{figure}

In this section we investigate of fate of scarring when quenching from the extreme vacua to finite values of $\mu$ and $g^2$. As in the main text, to probe this we study the difference between the maximal and minimal fidelity revival after a quench. The results can be seen in Fig.~\ref{fig:LGT_quench_0m}, where the most striking feature for each $S$ is a main diagonal (red dashed) line showing close to perfect revivals.
This line corresponds to $\mu=(2S-1)g^2/4$. Indeed, at that point and for $\mu,g^2\gg J/\sqrt{S}$, the Hilbert space fractures and the $\ket{0_-}$ state and its translated counterpart both find themselves in a fragment of the Hilbert space that can be mapped to free spin-$1/2$ paramagnets with $L/2$ spins. We now demonstrate this by starting in $\ket{0_-}=\ket{S,-S,S,-S,\ldots}$. The total energy of that state is equal to $Lg^2S^2/2$. At first, the only move allowed in the constrained Hilbert space is taking any of the sites with spin eigenvalue $S$ to $S-1$. This also conserves the total energy. Indeed, this operation creates two fermions that lead to an increase in energy of $2\mu=g^2(2S-1)/2$, thus counteracting the decrease of $g^2[S^2-(S-1)^2]/2=g^2(2S-1)/2$ in electromagnetic energy. Hence, every other site can freely flip between $S$ and $S-1$. However, the next allowed step would be to take a site with spin eigenvalue $-S$ to $1-S$. This leads to a large energy change as it both destroys fermions and reduces the electromagnetic energy. If this energy change is much larger than the hopping energy, which is roughly equal to $J/\sqrt{S}$ near the end of the chain, this move is greatly suppressed, hence these sites are frozen at $-S$. We thus have a system that is equivalent to a set of $L/2$ noninteracting two-level systems.
The same is true when starting in the state $\ket{-S,S,-S,S,\ldots}$, but the odd and even sites are now flipped.
It is important to distinguish these perfect revivals from scarring, as they emerge here as a sole consequence of Hilbert space fragmentation. Every product state located entirely within one of these two noninteracting disconnected subspaces will show perfect revivals with the same frequency.

\section{Quantum dimension}\label{app:QD}

We now derive the quantum dimensions for the TSM/QLM and the generalized PXP model. By grouping states in equivalence classes depending on the value of their leftmost site, it is possible to get a transfer matrix that relates the class sizes for system sizes $L$ and $L-1$. The quantum dimension is then simply the dominant eigenvalue of this transfer matrix.

For the QLM (and equivalently, the TSM, as the constraint is the same for both), we simply need to keep track of the state of the leftmost site and the transfer matrix is of size $(2S+1) \times (2S+1)$. We can order the spin values by ``compatibility'', which means for integer spin $0,-1,1,-2,2,\ldots,-S,S$ and for half-integer $-1/2,1/2,-3/2,3/2,\ldots,-S,S$. The advantage of this ordering is that, due to Gauss's law, a value is only compatible with the one before or after it, except the first one which is also compatible with itself. This allows us to write the transfer matrix as
\begin{equation}
    \mathcal{T}_\mathrm{QLM}(S)=\begin{pmatrix}1 & 1 & 0& 0\\ 1 & 0& \ddots & 0 \\ 0 & \ddots & 0 & 1 \\
0& 0 & 1 & 0 \end{pmatrix},
\end{equation}
which has the form of a tridiagonal matrix with unity entries on the super- and subdiagonal, and zeros everywhere else except at the $(1,1)$ entry, which also carries a value of $1$. The number of states in the case of PBC is given by
\begin{equation}
\mathcal{D}_\mathrm{QLM}(S,L){=}{\rm Tr}\left[(\mathcal{T}_\mathrm{ QLM}(S))^L\right]\hspace{-0.1cm}{=}\hspace{-0.17cm}\sum_{j=1}^{2S+1}\lambda_j^L {\underset{L\to\infty }{\rightarrow}}\lambda_{2S+1}^L,
\end{equation}
where the $\lambda_j$ are the eigenvalues of $\mathcal{T}_\mathrm{ QLM}(S)$ sorted by increasing modulus. For large $L$, the dominant eigenvalue is the only one contributing significantly and so its value sets the quantum dimension.  
Let $d=2S+1$, then the characteristic polynomial of the transfer matrix is
\begin{equation}
P_\mathrm{QLM}(S,\lambda)=\lambda^{d}+\sum_{i=1}^{d}(-1)^{\lceil i/2 \rceil}\lambda^{d-i}\binom{d-\lceil i/2 \rceil}{\lfloor i/2 \rfloor},
\end{equation}
where the $\binom{n}{k}$ are the binomial coefficients. For example,
$P_{S=1/2}=\lambda^2-\lambda\binom{1}{0}-\binom{1}{1}=\lambda^2-\lambda-1$ and $P_{S=1}=\lambda^3-\lambda^2\binom{2}{0}-\lambda\binom{2}{1}+\binom{1}{1}=\lambda^3-\lambda^2-2\lambda^2+1$. The quantum dimension is the largest root of this polynomial, which is given by Eq.~\eqref{eq:QD_TSM} in the main text, and which clearly becomes $2$ at $S\to\infty$. Table \ref{tab:QD} in the main text shows its values for selected values of $S$. Note that in Ref.~\cite{Zache2021achieving}, this quantum dimension was also calculated as
\begin{align}
    \alpha_{S,L}=2\Bigg[\sum_{j=1}^{2S+1}\cos^L\left(\frac{j\pi}{4S+3}\right)\Bigg]^{\frac{1}{L}},
\end{align}
for finite systems, which in the thermodynamic limit $L\to\infty$ converges to our result.

For the generalized PXP model~\cite{wenwei18TDVPscar}, the procedure is even simpler. In that case, we can split the states into two equivalence classes: states with the leftmost site equal to $-S$ and other states. The resulting transfer matrix is 
\begin{equation}
    \mathcal{T}_\mathrm{PXP}(S)=\begin{pmatrix}1 & 2S \\ 1 & 0 \end{pmatrix}.
\end{equation}
The characteristic polynomial is easy to compute as $P_\mathrm{PXP}(S,\lambda)=\lambda^2-\lambda-2S$, giving a dominant eigenvalue of $\lambda=(1+\sqrt{1+8S})/2$, matching the result in Ref.~\cite{wenwei18TDVPscar}.

\section{Detailed study of the $S=1$ case}\label{app:S1_detailed}

The case of $S=1$ in the TSM and QLM is the simplest example that is different from the PXP model \eqref{eq:PXP}. Here, we show that the other approaches developed to investigate the PXP model and enhance its revivals also work for the spin-$1$ $\mathrm{U}(1)$ QLM and TSM, which are identical to each other for $S=1$.

For $S=1$, the Hamiltonians in Eqs.~\eqref{eq:H_spin} and \eqref{eq:H_TSM} are equal and can be written with local constraints as
\begin{align}\nonumber
    \hat{H}&=\frac{J}{\sqrt{8}}\,\sum_{j=1}^L\Big[\hat{\mathcal{P}}^{-1}_{j-1}\left( \hat{s}^+_j\hat{\mathcal{P}}^0_j\right)\hat{\mathcal{P}}^{-1}_{j+1}+\\\nonumber
    &+\hat{\mathcal{P}}^{0}_{j-1}\left(\hat{\mathcal{P}}^0_j\hat{s}^+_j\right)\hat{\mathcal{P}}^{0}_{j+1}+\mathrm{H.c.}\Big]\\\label{eq:H_S1}
    &-2\mu\sum_j \hat{s}^z_j+\frac{g^2}{2}\sum_j\left(\hat{s}_j^z\right)^2,
\end{align}
where $\hat{\mathcal{P}}^M$ is the projector on the state with spin eigenvalue $M$.

\begin{figure}[t!]
\centering
\includegraphics[width=\linewidth]{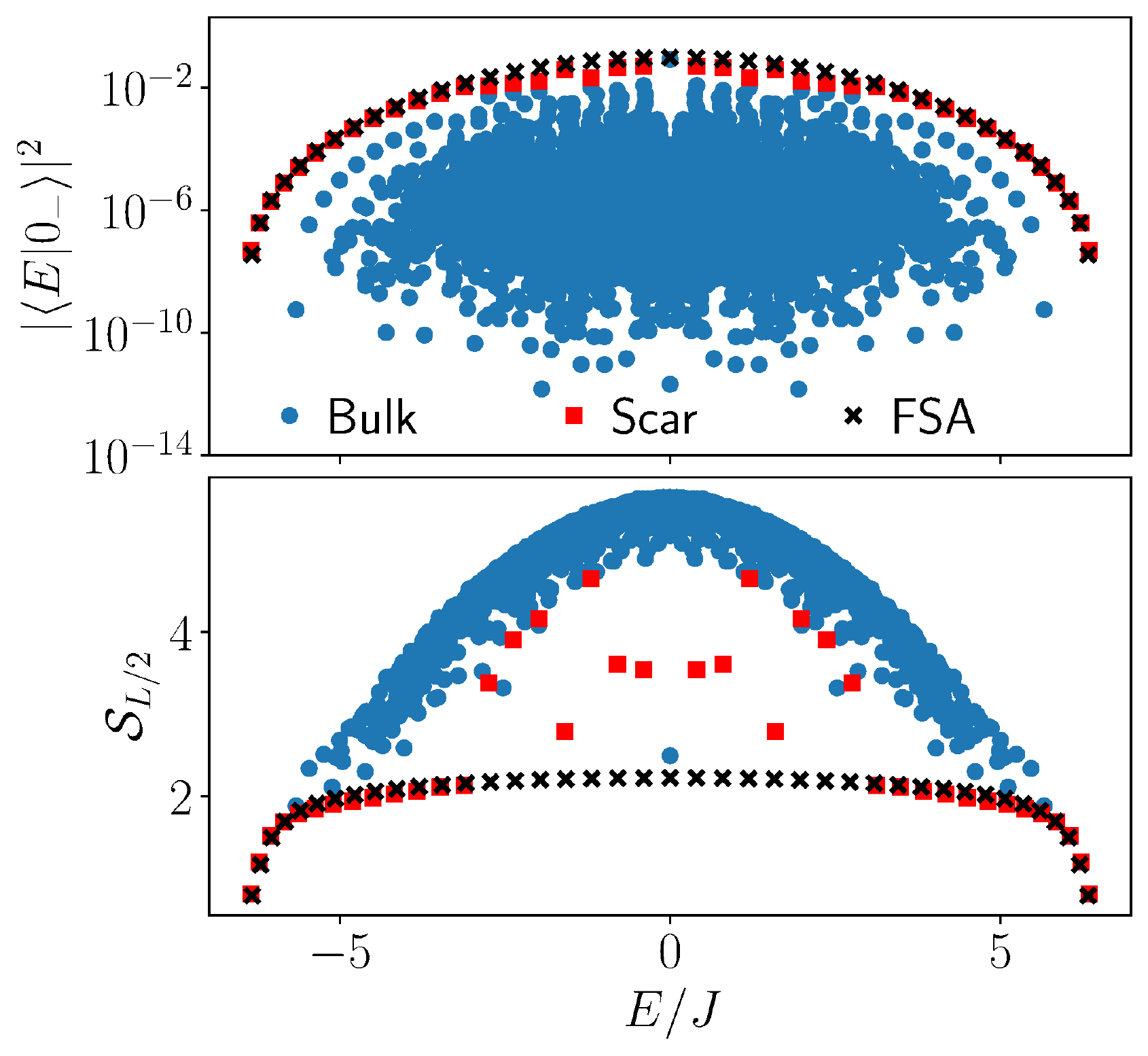}
\caption{Overlap of the $\ket{0_-}$ state with the eigenstates (top) and half-chain entanglement entropy (bottom) in the QLM/TSM with $S=1$ and $L=20$. The FSA results closely reproduce the exact diagonalization results, in particular for the overlaps with $\ket{0_-}$. In contrast, the entanglement entropy of scarred eigenstates in the middle of the spectrum is less accurately reproduced, due to the strong mixing of states within the towers.
}
\label{fig:LGT_S1}
\end{figure}

\subsection{Forward scattering approximation}
As in the case of $S=1/2$, for $S=1$ we can use the Forward Scattering approximation (FSA) to approximate the eigenstates \cite{Turner2018}.
First, we need to decompose the off-diagonal part of the Hamiltonian into raising and lowering operators $\hat{H}^+$ and $\hat{H}^-$, such that $\hat{H}^+$ takes us away from the initial reviving state $\ket{0_-}=\ket{1,-1,1,-1,\ldots,1,-1}$ while $\hat{H}^-=\big(\hat{H}^+\big)^\dagger$. It is fairly straightforward to derive the exact form of these expressions as
\begin{align}\nonumber
    \hat{H}^+&{=}\frac{1}{\sqrt{8}}\sum_{j=1}^{L/2}\Big[\hat{\mathcal{P}}^0_{2j-1}\left(\hat{s}^-_{2j}\hat{\mathcal{P}}^0_{2j} \right)\hat{\mathcal{P}}^0_{2j+1}\\\nonumber
    &{-}\hat{\mathcal{P}}^-_{2j-1}\left(\hat{\mathcal{P}}^0_{2j}\hat{s}^-_{2j} \right)\hat{\mathcal{P}}^-_{2j+1} \\\nonumber
    &+\hat{\mathcal{P}}^0_{2j-2}\left(\hat{\mathcal{P}}^0_{2j-1}\hat{s}^+_{2j-1} \right)\hat{\mathcal{P}}^0_{2j}\\
    &{-}\hat{\mathcal{P}}^-_{2j}\left(\hat{s}^+_{2j-1}\hat{\mathcal{P}}^0_{2j-1} \right)\hat{\mathcal{P}}^-_{2j}\Big].
\end{align}
Using this raising operator and starting from the state $\ket{F_0}=\ket{0_-}=\ket{1,-1,1,-1, \ldots,1,-1}$, we can build the FSA states $\ket{F_n}=\frac{1}{\mathcal{N}}\big(\hat{H}^+\big)^n\ket{F_0}$ for $n=0$ to $2L$, with $\mathcal{N}$ denoting a normalization factor. It is useful to note that $\ket{F_{2L}}=\ket{-1,1,-1,1\ldots -1,1}$ and $\hat{H}^+\ket{F_{2L}}=0$. We can project the Hamiltonian to this set of $2L+1$ states, which leads us to a tridiagonal matrix with off-diagonal elements $\beta_{n,n+1}$ as $\bra{F_m}\hat{H}\ket{F_n}=\beta_{n,n+1}\delta_{m,n+1}+\beta^*_{n-1,n}\delta_{m,n-1}$. As for the spin-$1/2$ PXP model, the eigenstates in this low-dimensional subspace have energies very close to the ones of the actual scarred eigenstates \cite{Turner2018}. This is demonstrated in Fig.~\ref{fig:LGT_S1}. The main difference between the FSA states and the exact scarred eigenstates lies in their number. Indeed, in the FSA we obtain exactly $2SL+1$ states, whereas in reality the model has $2SL+1$ towers of states. So the FSA only gives one scarred state per tower. As a result, the FSA eigenstates are very atypical, having a very low entanglement entropy. This atypicality is in fact ``diluted'' amongst the many eigenstates in each tower, leading to individual states with larger entanglement entropy.

\begin{figure}[t!]
\centering
\includegraphics[width=\linewidth]{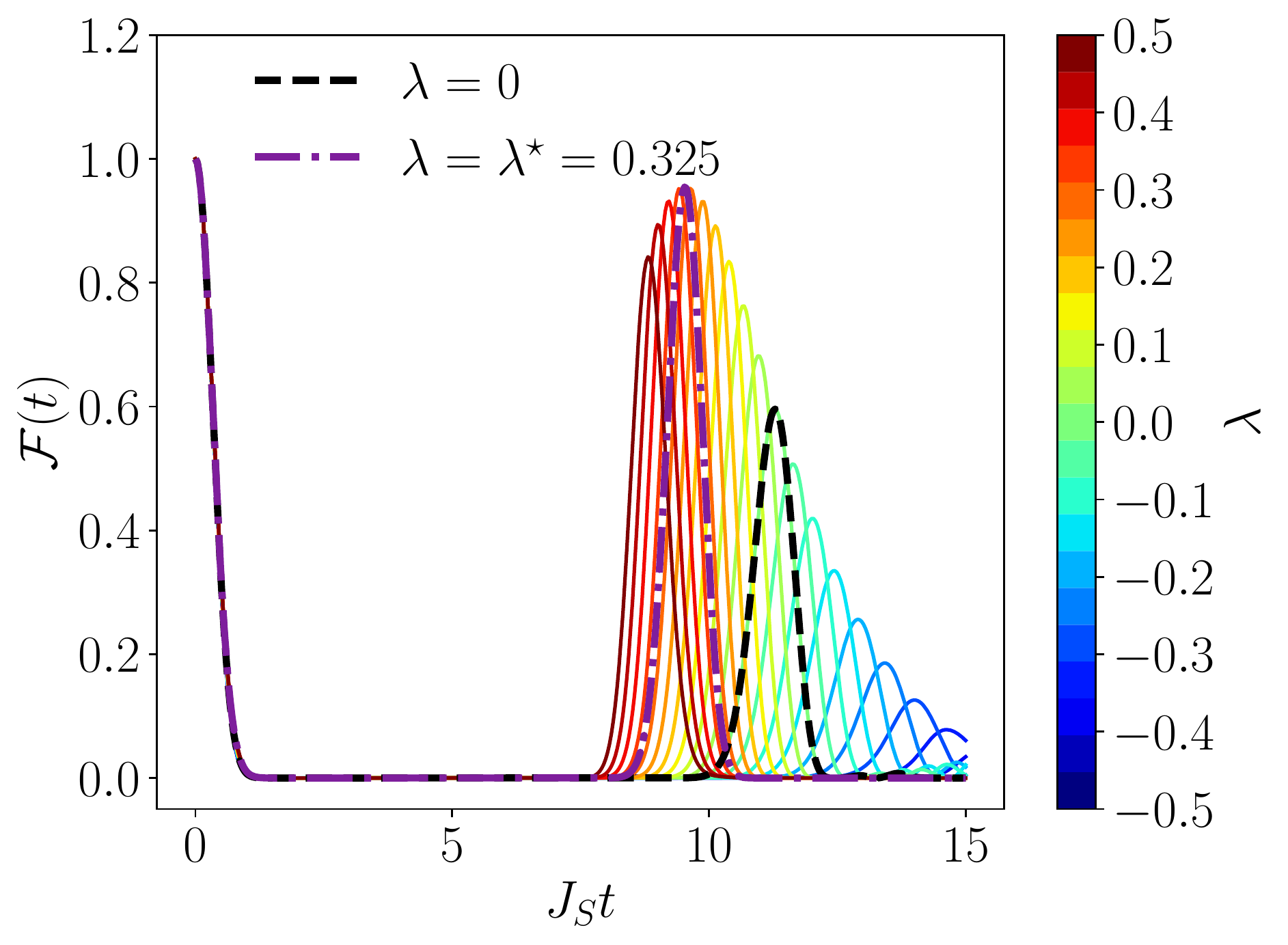}
\caption{Revivals in the QLM/TSM for $S=1$ and $L=18$ with different strengths of the perturbation $\delta \hat{H}_{(1)}$. For $\lambda=0.325$, the revivals become close to perfect.}
\label{fig:LGT_S1_pert}
\end{figure}  

\subsection{Algebra-correcting perturbation for $S=1$}\label{app:pert}

As for the PXP model with spins-$1/2$~\cite{Choi2018}, the accuracy of the FSA implies that there is an approximate $\mathrm{SU}(2)$ algebra structure in the scarred subspace. We can derive a perturbation to ``correct'' this algebraic structure by following the prescription in Ref.~\cite{Bull2020}. We consider $\hat{H}^+$ and $\hat{H}^-$ to be the raising and lowering operators of that algebra. We can derive the effective $\hat{H}^z$ operators as 
\begin{align}\nonumber
    \hat{H}^z&=2[\hat{H}^+,\hat{H}^-]\\\nonumber
    &=\frac{1}{\sqrt{8}}\sum_{j=1}^{L/2}\Big[ \hat{\mathcal{P}}^0_{2j-1}\left(\hat{\mathcal{P}}^0_{2j}-\hat{\mathcal{P}}^{-1}_{2j}\right)\hat{\mathcal{P}}^0_{2j+1} \\\nonumber
    &+ \hat{\mathcal{P}}^{-1}_{2j-1}\left(\hat{\mathcal{P}}^1_{2j}-\hat{\mathcal{P}}^{0}_{2j}\right)\hat{\mathcal{P}}^{-1}_{2j+1} \\\nonumber
    &-\hat{\mathcal{P}}^0_{2j-2}\left(\hat{\mathcal{P}}^0_{2j-1}-\hat{\mathcal{P}}^{-1}_{2j-1}\right)\hat{\mathcal{P}}^0_{2j} \\
    &- \hat{\mathcal{P}}^{-1}_{2j}\left(\hat{\mathcal{P}}^1_{2j-1}-\hat{\mathcal{P}}^{0}_{2j-1}\right)\hat{\mathcal{P}}^{-1}_{2j} \Big].
\end{align}
From there, we compute the commutators 
\begin{subequations}
\begin{align}
    [\hat{H}^z,\hat{H}^+]&=\hat{H}^++\hat{\delta}^+_{1},\\
    [\hat{H}^z,\hat{H}^-]&=-\hat{H}^-+\hat{\delta}^-_{1},
\end{align}
\end{subequations}
which obey the expected $\mathrm{SU}(2)$ commutation rules, up to error terms denoted by $\hat{\delta}_1^\pm$. The latter are given by
\begin{align}\nonumber
\hat{\delta}_{(1)}^+=&-\frac{1}{4\sqrt {2}}\sum_{j=1}^{L/2}\Big[ 2\hat{\mathcal{P}}^0_{2j-2}\left(\hat{s}^-_{2j-1}\hat{\mathcal{P}}^0_{2j-1}\right)\hat{\mathcal{P}}^0_{2j}\\\nonumber
&+2\hat{\mathcal{P}}^0_{2j-1}\left(\hat{\mathcal{P}}^0_{2j}\hat{s}^+_{2j}\right)\hat{\mathcal{P}}^0_{2j+1}\\\nonumber
&+\hat{\mathcal{P}}^{-1}_{2j-1}\left(\hat{s}^+_{2j}\hat{\mathcal{P}}^0_{2j}\right)\hat{\mathcal{P}}^{-1}_{2j+1}\hat{\mathcal{P}}^0_{2j+2}\\\nonumber
&+\hat{\mathcal{P}}^0_{2j-2}\hat{\mathcal{P}}^{-1}_{2j-1}\left(\hat{s}^+_{2j}\hat{\mathcal{P}}^0_{2j}\right)\hat{\mathcal{P}}^{-1}_{2j+1}\\\nonumber
&+\hat{\mathcal{P}}^{-1}_{2j-2}\left(\hat{\mathcal{P}}^0_{2j-1}\hat{s}^-_{2j-1}\right)\hat{\mathcal{P}}^{-1}_{2j}\hat{\mathcal{P}}^0_{2j+1}\\
&+\hat{\mathcal{P}}^0_{2j-3}\hat{\mathcal{P}}^{-1}_{2j-2}\left(\hat{\mathcal{P}}^0_{2j-1}\hat{s}^-_{2j-1}\right)\hat{\mathcal{P}}^{-1}_{2j} \Big],
\end{align}
and $\hat{\delta}^-_{(1)}=-\big[\hat{\delta}^+_{(1)}\big]^\dagger$.
We can then partially cancel the unwanted error terms by introducing the following perturbation to the model
\begin{align}\nonumber
    \delta \hat{H}_{(1)}&=\hat{\delta}^-_{(1)}-\hat{\delta}^+_{(1)}\\\nonumber
    &=\frac{1}{4\sqrt {2}}\sum_{j+1}^L \Big[2\hat{\mathcal{P}}^0_{j-1}\left(\hat{\mathcal{P}}^0_j\hat{s}^+_j\right)\hat{\mathcal{P}}^0_{j+1}\\\nonumber
    &+\hat{\mathcal{P}}^{-1}_{j-1}\left(\hat{s}^+_j\hat{\mathcal{P}}^0_j\right)\hat{\mathcal{P}}^{-1}_{j+1}\hat{\mathcal{P}}^0_{j+2} \\
    &+\hat{\mathcal{P}}^0_{j-2}\hat{\mathcal{P}}^{-1}_{j-1}\left(\hat{s}^+_j\hat{\mathcal{P}}^0_j\right)\hat{\mathcal{P}}^{-1}_{j+1} +\mathrm{H.c.}\Big].
\end{align}
We add this perturbation to the Hamiltonian, $\hat H \to \hat H + \lambda \cdot \delta \hat H_{(1)}$, and look for the value $\lambda$ that gives the best revivals for the extreme vacua. From Fig.~\ref{fig:LGT_S1_pert} we see that we can get a substantial improvement of the revivals for $\lambda \approx 0.325$, with the revivals being close to perfect in that case.
This procedure can be carried out in the same way for arbitrary spin length $S$. However, the number of local terms in $\hat{H}^+$ increases linearly with $S$ and the derivation becomes increasingly tedious.

\section{Bond dimension of the constraint for the QLM and TSM}\label{app:MPO}
We argue that to represent the constraint in the QLM and TSM model as a matrix-product operator (MPO), we need a bond dimension scaling as $2S+1$. This is in stark contrast with the generalized PXP model where the bond dimension is equal to $2$ for any $S$ \cite{wenwei18TDVPscar}.

The first step is to write the global constraint $\mathcal{P}$ as a sum of products of local two-site constraints $\mathcal{P}=\prod_i\mathcal{P}_{i,i+1}$. Each local constraint can be expressed as
\begin{equation}
\mathcal{P}_{i,i+1}{=}\hat{P}_i^{-S}\hat{P}_{i+1}^{S}+\hspace{-0.25cm}\sum_{m=1-S}^{S}\left(\hat{P}_i^{m}\hat{P}_{i+1}^{-m}+\hat{P}_i^{m}\hat{P}_{i+1}^{-m-1}\right),
\end{equation}
where $\hat{P}^m_i=\ket{m}\bra{m}$ is the projector on the spin eigenstate with magnetization $m$ along the $z$-direction. Already, this sum of $4S$ local terms contrasts with the PXP constraint which can be written as
$\hat{P}_i^{-S}\hat{P}_{i+1}^{-S}+\hat{Q}_i\hat{P}_{i+1}^{-S}+\hat{P}_i^{-S}\hat{Q}_{i+1}$ for any $S$, with $\hat{Q}=\mathds{1}-\hat{P}^{-S}$. 

\begin{figure}[t!]
\centering
\includegraphics[width=\linewidth]{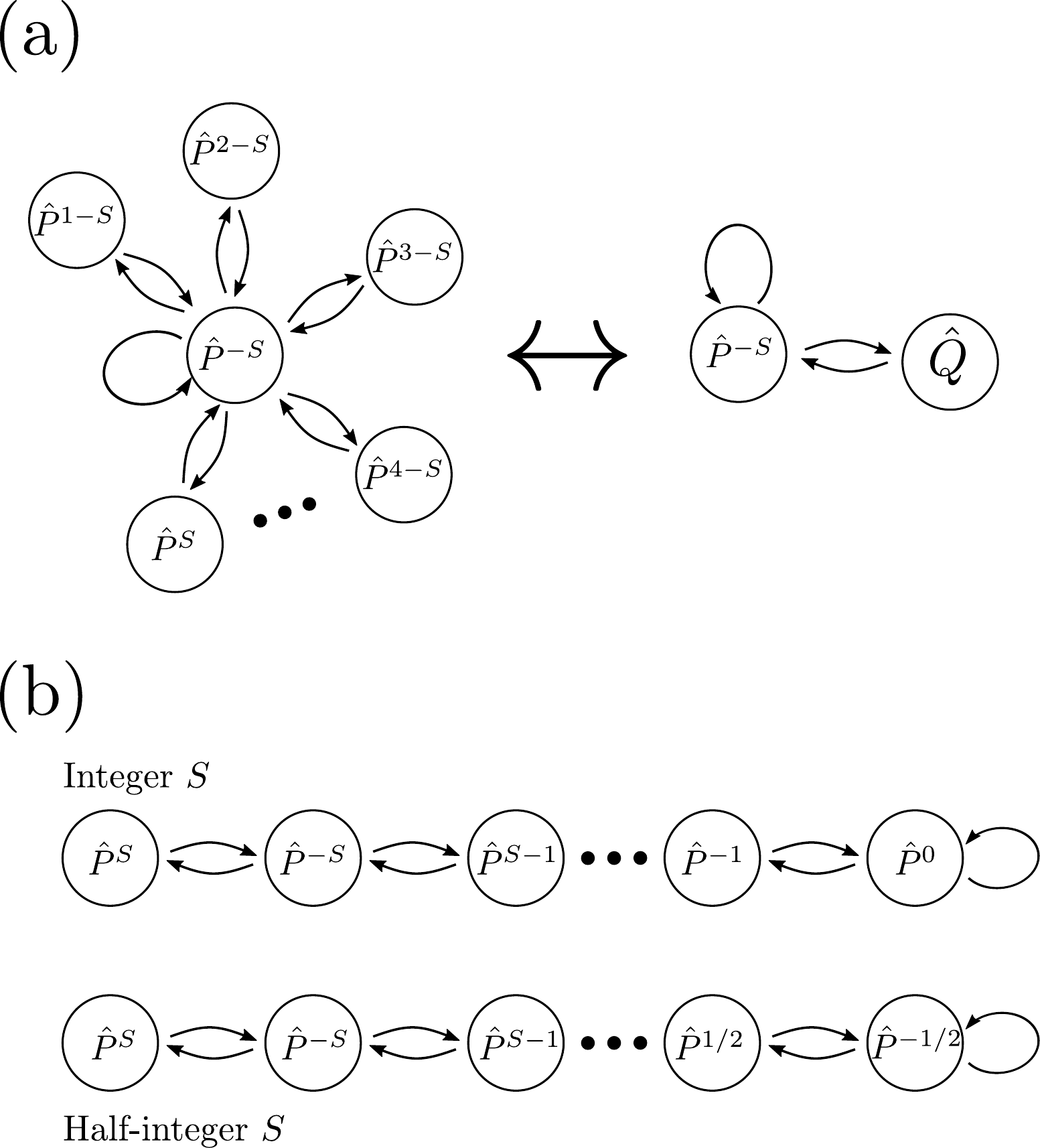}
\caption{State machine representing the local two-site constraint for (a) the generalized PXP model and (b) the QLM and TSM. For the PXP model, the state machine can be easily reduced to a total of two states by grouping together all states having the same rules. This is not the case for the QLM and TSM as all states have different rules.}
\label{fig:state_machine}
\end{figure}

To find the minimal bond dimension, we need to represent the constraint as an MPO. A common approach is to write it as a state machine \cite{mps2,CrosswhiteMPO}, where each transition must be between operators that can be placed next to each other. This is straightforward and the result is shown in Fig.~\ref{fig:state_machine}.
While both states can be written to have $2S+1$ states, one can see that for the PXP model all states with $m\neq-S$ have exactly the same rules and so can be grouped together. For the QLM, all states have different rules and so there is no simple way to merge them. From the state machine one can then write the MPO as an operator-valued matrix where the element $(i,j)$ is equal to $\hat{P}^{-S+i-1}$ if the transition $\hat{P}^{-S+i-1}$ to $\hat{P}^{-S+j-1}$ is allowed and $0$ otherwise.   
The resulting MPO are then,
\begin{equation}
\begin{aligned}
\hat{W}_\mathrm{PXP}&=\begin{pmatrix}\hat{P}^{-S} & \hat{P}^{1-S} & \ldots & \hat{P}^{S}\\ \hat{P}^{-S} & 0& 0 & 0 \\ \vdots & 0 & 0 &0 \\
\hat{P}^{-S}& 0 & 0 & 0 \end{pmatrix}\rightarrow \begin{pmatrix}\hat{P}^{-s} & \hat{Q} \\ \hat{P}^{-s} & 0 \end{pmatrix}, \\ 
\hat{W}_\mathrm{QLM}&=\begin{pmatrix}\hat{P}^{0} & \hat{P}^{-1} & 0& 0\\ \hat{P}^{0} & 0& \ddots & 0 \\ 0 & \ddots & 0 &\hat{P}^S \\
0& 0 & \hat{P}^{-s} & 0 \end{pmatrix},
\end{aligned}
\end{equation}
where we assumed integer spin for the QLM. For half-integer spin we obtain the same structure but with $\hat{P}^{-1/2}$ in the leftmost column instead of $\hat{P}^0$. These matrices have exactly the same structure as the transfer matrices used in Appendix~\ref{app:QD}. This is of course not a coincidence, as they encode the same constraint. However, in the present case they are operator-valued matrices.

It is important to note here that for the PXP model, even if we write it with all $\hat{P}$ operators explicitly, the MPO always has rank 2. For the QLM, the matrix has full rank  $2S+1$. While it is still possible that the bond dimension could be reduced using a nontrivial global gauge transformation, the simple recipe used for PXP is not valid for that model and it is likely that the bond dimension cannot be reduced below $2S+1$. This would make any analytical TDVP approach quite cumbersome to work with for larger $S$ in the case of the TSM/QLM.

\section{Symmetric subspace for higher $S$}\label{app:sym_subs}

Here, we provide further details about the generalization of the symmetric subspace introduced in Ref.~\cite{TurnerSym} to constrained models with higher spin. We develop two different subspaces $\mathcal{K}_2$ and $\mathcal{K}_{4S}$, where the subscript denotes the number of values characterizing each basis state in the subspace. 

For $\mathcal{K}_2$,  we first group all states based on the total number of excitations on each sublattice (encompassing all odd and even sites respectively). By `excitations', we mean the number of times the spin raising operator was applied on top of the lowest spin states. For example, we can consider the case of $S=3/2$ and the state $\ket{-1/2,-3/2,3/2,1/2}$. On the first sublattice we need one application of $\hat{s}^+$ to go from $-3/2$ to $-1/2$ and three applications to go from $-3/2$ to $3/2$. Hence, there are $3+1=4$ excitations on it. On the second sublattice, there are $0+2=2$ excitations. This state will then belong to the equivalence class $(n_1=4,n_2=2)$. Each $n_j$ takes values  between 0 and $SL$, and so the number of classes scales as $\mathcal{O}\left(S^2L^2\right)$, which is much slower than the full Hilbert space.

We can then define the basis of $\mathcal{K}_2$ by creating one basis state per equivalence class. Following the TDVP Ansatz in \cite{wenwei18TDVPscar}, we want our basis states to span a subspace formed by coherent spin states that do not violate the constraint on the Hilbert space. For the latter part, this means simply discarding any state that would violate the constraint. For the former, we have to make sure that our basis states can reproduce the spin coherent states defined on a single site as
\begin{equation}
\begin{aligned}
&\ket{(\theta,\phi)}=e^{i\phi (\hat{s}^z+S)}e^{-i\theta \hat{s}^x}\ket{-S} \\
&=\hspace{-0.11cm}\sum_{k{=}0}^{2S}\hspace{-0.1cm} \sqrt{\binom{2S}{k}} e^{ik(\phi-\pi/2)} \cos^{2S{-}k}\left(\frac{\theta}{2}\right)\sin^{k}\left(\frac{\theta}{2}\right)\ket{k-S}.
\end{aligned}
\end{equation}
The main prefactor to look for here is the square root of the binomial coefficient. Indeed, if we now consider a state with spin coherent states on multiple sites we have to multiply all prefactors. So on a single sublattice, if we have a total of $n$ excitations spread onto $k$ sites, we will end up with a total prefactor of
\begin{equation}\label{eq:perm_coher}
    e^{in(\phi-\pi/2)}\cos^{2kS{-}n}\left(\frac{\theta}{2}\right)\sin^{n}\left(\frac{\theta}{2}\right)\prod_{j=1}^k \sqrt{\binom{2S}{S{-}m_j}},
\end{equation}
where we could remove the sine, cosine, and exponential terms from the product because their final exponent only depends on the total number of excitations, which we know is equal to $nS$. Since these three prefactors are the same for all states in an equivalence class, they do not need to be incorporated into the basis. However, this is not the case for the binomial coefficient, as those vary from state to state. Hence, the basis states of our subspace $\mathcal{K}_2$ take the form
\begin{equation}
    \ket{n_1,n_2}=\frac{1}{\mathcal{N}}\sum_{\ket{\phi}\in(n_1,n_2)}\left(\prod_{j=1}^L \sqrt{\binom{2S}{S+m_j}} \right)\ket{\phi},
\end{equation}
where the sum is over all basis states in the equivalence class $(n_1,n_2)$, $\mathcal{N}$ is a normalization factor and the $m_j$ are the spin eigenvalues of the individual sites for the sate $\ket{\phi}$. If there were no constraint, this would just be the set of \emph{global} spin coherent states on each sublattice. However, as the equivalence classes only contain states that do not violate the constraint, they are nontrivial. 

We now give the details on how to construct the basis of the larger subspace $\mathcal{K}_{4S}$. The goal of this subspace is to describe the state of the big spin formed by each of the $4S$ hypercubes in the QLM graph. Each of them corresponds to having one sublattice with all sites at a fixed spin eigenvalue $m$, and the other sublattice flipping freely between $-m$ and $-m-1$. As a consequence we have to keep track of the number of sites having the spin eigenvalue $m$ on each sublattice for all $m$. As such, for each sublattice we will have $2S$ numbers characterizing each state, corresponding to the number of sites with spin eigenvalue $m$. Here we only deal with hypercubes, corresponding to sets of spins-$1/2$, and we do not need to incorporate the prefactor as for the coherent states with higher spin in $\mathcal{K}_2$. So each basis state is simply a symmetric superposition of all states with the same population distribution (the same number of sites with spin eigenvalue $m$ for each $m$) in both sublattices. For example, for $S=1$ and $L=4$ the state $\ket{n_1^0=1,n_1^1=2,n_2^0=0,n_2^1=2}$ would be defined as 
\begin{align}\nonumber
    \ket{n_1^0{=}1,n_1^1{=}0,n_2^0{=}0,n_2^1{=}2}{=}&\frac{1}{\sqrt{2}}\big(\ket{0,1,{-}1,1}\\
    &+\ket{{-}1,1,0,1}\big).
\end{align}
Note that $\mathcal{K}_{4S}=\mathcal{K}_{2}$ for $S=1/2$ , and for higher $S$ we have $\mathcal{K}_{2} \subset \mathcal{K}_{4S}$. Indeed, as the population distribution in every basis state of $\mathcal{K}_{4S}$ is the same, all of the states in it would get the same prefactor in Eq.~\eqref{eq:perm_coher}. As such, every single basis state of $\mathcal{K}_{2}$ is in $\mathcal{K}_{4S}$, while the opposite is not true.

\bibliography{Schwinger_biblio}

\end{document}